\documentclass[prd,aps,floats,floatfix,eqsecnum,nofootinbib]{revtex4}
\usepackage{verbatim,graphicx,amssymb,amsbsy,bm,amsmath,rotating,epsfig}
\usepackage{psfrag}
\usepackage{hyperref}
\usepackage{fontenc}
\newcommand{\be}{\begin{equation}}
\newcommand{\ee}{\end{equation}}
\newcommand{\bea}{\begin{eqnarray}}
\newcommand{\eea}{\end{eqnarray}}

\newcommand{\vx}{\ensuremath{\vec{x}}}
\newcommand{\va}{\ensuremath{{\vec \alpha}}}
\newcommand{\vk}{\ensuremath{\vec{k}}}
\newcommand{\vka}{\ensuremath{\vec{\kappa}}}
\newcommand{\vq}{\ensuremath{\vec{q}}}
\newcommand{\cq}{\ensuremath{\check{q}}}
\newcommand{\cQ}{\ensuremath{\check{Q}}}
\newcommand{\vQ}{\ensuremath{\vec{Q}}}

\begin{document}
\title{Cosmological evolution of warm dark matter fluctuations I: 
Efficient computational framework with Volterra integral equations}
\author{\bf H. J. de Vega $^{(a,b)}$}
\email{devega@lpthe.jussieu.fr} 
\author{\bf N. G. Sanchez $^{(b)}$}
\email{Norma.Sanchez@obspm.fr} 
\affiliation{$^{(a)}$ LPTHE, Universit\'e
Pierre et Marie Curie (Paris VI) et Denis Diderot (Paris VII),
Laboratoire Associ\'e au CNRS UMR 7589, Tour 13-14, 4\`eme. et 5\`eme. \'etages, 
Boite 126, 4, Place Jussieu, 75252 Paris, Cedex 05, France. \\
$^{(b)}$ Observatoire de Paris,
LERMA, Laboratoire Associ\'e au CNRS UMR 8112.
 \\61, Avenue de l'Observatoire, 75014 Paris, France.}
\date{\today}
\begin{abstract}
We study the complete cosmological evolution of dark matter (DM) density fluctuations for 
DM particles that decoupled being ultrarelativistic during the radiation 
dominated era which is the case of keV scale warm DM (WDM). 
The new framework presented here can be applied to 
other types of DM and in particular we extend it to cold DM (CDM). 
The collisionless and linearized Boltzmann-Vlasov equations (B-V) 
for WDM and neutrinos in the presence of photons and coupled to the 
linearized Einstein equations are studied in detail in the presence of anisotropic 
stress with the Newtonian potential generically different from the spatial curvature 
perturbations. We recast this full system of B-V equations for DM and neutrinos into 
a system of coupled Volterra integral equations.
These Volterra-type equations are valid both in the 
radiation dominated (RD) and matter dominated (MD) eras during which
the WDM particles are ultrarelativistic and then nonrelativistic.
This generalizes the so-called Gilbert integral equation only valid for 
nonrelativistic particles in the MD era. We succeed to reduce the system of 
four Volterra integral equations for the density and anisotropic stress 
fluctuations of DM and neutrinos into a system of only two coupled Volterra 
equations. The kernels and inhomogeneities in these equations are
explicitly given functions. 
Combining the Boltzmann-Vlasov equations and the linearized 
Einstein equations constrain the initial conditions on the distribution
functions and gravitational potentials. In the absence of neutrinos 
the anisotropic stress vanishes and the Volterra-type equations reduce 
to a single integral equation. These Volterra integral equations 
provide a useful and precise framework to compute the primordial WDM 
fluctuations over a wide range of scales including small scales up to 
$ k \sim 1/5 \; {\rm kpc} $.

\end{abstract}
\maketitle
\tableofcontents

\section{Introduction and Summary of Results}

The evolution of the dark matter (DM) density fluctuations since the DM decoupling 
till today is a basic problem in cosmology. This problem has been extensively treated in the literature 
for particles decoupling being nonrelativistic (Cold DM, CDM) \cite{dod,mab,ws,sz,gor}. 

\medskip

Particles decoupling ultrarelativistically in the radiation dominated era 
(Warm DM, WDM) were proposed as DM candidates 
years ago \cite{primeros,pb,dw,neus}. Such WDM particles with mass in the keV scale become the subject 
of a renewed interest in recent years \cite{mas1,bdvs2,dvs,dvs2,neus2,mas2,mas3}. 

\medskip

In this paper, we study the evolution of DM density fluctuations for particles
that decoupled being ultrarelativistic during the radiation dominated era.
(Ref. \cite{bwu} has recently considered this issue).

\medskip

The expansion of the Universe dilutes matter in the early universe
and particle decoupling happens when the particle collisions become 
sufficiently rare and can be neglected. Therefore, and it is well known,
the particle distribution generically freezes out at decoupling.
This happens irrespective of whether the particles are in or out of thermal 
equilibrium (see ref. \cite{gor}, section 2 of ref.\cite{bdvs2} and ref. 
\cite{weinb})

The treatment of the cosmology density fluctuations
presented here and in the companion paper ref. \cite{dos}
is valid for generic freezed out distribution functions,
whether at thermal equilibrium or out of thermal equilibrium
and holds irrespective of the particular DM particle model.

\medskip

The linearized Boltzmann-Vlasov (B-V) equation provides an appropriate 
framework to follow the evolution of the primordial fluctuations since 
the DM decoupling till today. The linearized B-V equation turns to be 
particularly difficult to solve since it is in general a partial 
differential equation on a distribution function which depends on seven 
variables. Two strategies have been used to solve the linearized B-V 
equation.
One method consists in expanding the distribution function on Legendre polynomials transforming
the B-V equation into an infinite hierarchy of coupled ordinary
differential equations (ODE) \cite{dod,mab,ws,sz}. 
Another approach to the linearized B-V equation
integrates the distribution function over the particle momenta and recast
the linearized B-V equation into a linear integral equation of Volterra type 
\cite{gil}-\cite{otros,weinb,primeros,bw,bdvs}. 
In the case of nonrelativistic particles in a matter dominated universe
this leads to the so called Gilbert equation \cite{gil}.
This approach leads to linear integral equations
of Volterra type while the Legendre polynomials expansion produces
an infinite hierarchy of coupled ODE. The Volterra type integral equation
exhibits a long-range memory of the gravitational interaction \cite{bdvs}.
However, the memory of the RD era turns out to fade out substantially in the MD era.

\medskip

In this paper we derive a system of integral equations of Volterra type
valid for relativistic as well as for non-relativistic particles 
propagating in the radiation and matter dominated eras. 
For warm dark matter and neutrinos we obtain a pair of coupled Volterra integral equations for
the density fluctuations and the anisotropic stress. 

\medskip

We start by writing down the collisionless Boltzmann-Vlasov equation
in a spatially flat FRW spacetime with adiabatic fluctuations in the conformal gauge. 
The distribution function $ {\tilde f}_{dm}(\eta, \vq, \vx ) $ of the DM particles after their
decoupling and to linear order in the fluctuations can be written as
\be\label{bollinI}
{\tilde f}_{dm}(\eta, \vq, \vx) = {\hat N_{dm}}\; g_{dm} \; {\hat f^{dm}}_0(q) + 
{\tilde f^{dm}}_1(\eta, \vq, \vx) = 
{\hat N_{dm}}\; {\hat f^{dm}}_0(q) \; g_{dm} \left[ 1 + {\tilde \Psi}_{dm}(\eta, \vq, \vx) \right] \; ,
\ee
where $ {\hat f^{dm}}_0(q) $ is the homogeneous and isotropic zeroth order 
distribution at decoupling, $ g_{dm} $ is the number of internal degrees of freedom of the DM particle
and $ {\hat N}_{dm} $ is a normalization factor. 
$ \eta $ is the conformal time, $ \vq $ and $ \vx $ stand for the particle 
momentum and position, respectively.
We use the superscript tilde in configuration space as $ {\tilde \Psi} (\vx) $
to indicate the Fourier transform of the momentum space function $ \Psi (\vk) $.
The superscript hat stands for dimensionful functions as $ {\hat f^{dm}}_0(q) $ whose dimensionless
counterpart $ f_0^{dm}(Q) $ does not bear a hat.

\medskip

Neutrinos are analogously described by a distribution function $ {\tilde f}_{\nu}(\eta, \vq, \vx ) $
\be\label{bolinuI}
{\tilde f}_{\nu}(\eta, \vq, \vx) = {\hat N_{\nu}}(\eta) \; g_{\nu} \; {\hat f}^{\nu}_0(q) 
+ {\tilde f}_1^{\nu}(\eta, \vq, \vx) = 
{\hat N_{\nu}}(\eta) \; {\hat f}^{\nu}_0(q) \; g_{\nu} \left[ 1 + 
{\tilde \Psi}_{\nu}(\eta, \vq, \vx) \right] \; ,
\ee
where $ {\hat f}^{\nu}_0(q) $ stands for the zeroth order Fermi-Dirac distribution function for neutrinos
after decoupling, $ g_{\nu} $ is the number of neutrino internal degrees of freedom and $ {\hat N_{\nu}}(\eta) $
is a normalization factor. 

We obtain as the collisionless B-V equation for DM including linear terms
in the fluctuations 
\be\label{bvI}
\frac{\partial{\tilde \Psi}_{dm}}{\partial \eta} + \frac1{E} \;
q_i \; \partial_i {\tilde\Psi}_{dm} + \frac{\partial \ln {\hat f^{dm}}_0}{\partial \ln q}
\left[\frac{\partial {\tilde \phi}}{\partial \eta} - \frac{E}{q^2} \; 
q_i \; \partial_i \; {\tilde \psi} \right] = 0 \; .
\ee
The neutrino distribution function obeys the massless version of eq.(\ref{bvI})
\be\label{bvnuI}
\frac{\partial{{\tilde\Psi}_{\nu}}}{\partial \eta} + 
n_i \; \partial_i \; {\tilde\Psi}_{\nu} + \frac{d\ln {\hat f}^{\nu}_0}{d\ln q}
\left[\frac{\partial  {\tilde\phi}}{\partial \eta} - n_i \; \partial_i \;  {\tilde\psi} \right] = 0  \; .
\ee
where $ {\tilde \psi} $ is the Newtonian potential
and $ {\tilde \phi} $ corresponds to the spatial curvature perturbation.

The B-V equations (\ref{bvI})-(\ref{bvnuI}) are coupled to the linearized 
Einstein equations for the gravitational potentials $ \tilde \psi $ and 
$ \tilde \phi $. After Fourier transforming, the linearized Einstein
equations read
\bea\label{Iecfidm}
&&3 \; h(\eta)  \; \frac{\partial \phi}{\partial \eta} + k^2 \; \phi(\eta, \vk) + 3 \; 
h^2(\eta) \; \psi(\eta, \vk) = -4 \, \pi \; G \; \left[\frac{
\Delta_{dm}(\eta,\vk) + \Delta_\nu(\eta,\vk)}{a^2(\eta)} + 4 \,  
a^2(\eta) \; \rho_\gamma(\eta) \; \Theta_0 (\eta, \vk) \right]  \; , \\ \cr 
&& \sigma(\eta, \vk) \equiv  \phi(\eta, \vk) - \psi(\eta, \vk)  = \frac{4 \, \pi \; G }{k^2 \; a^2(\eta)}
\; \left[\Sigma_{dm}(\eta, \vk)+\Sigma_\nu(\eta, \vk) \right] = \sigma_{dm}(\eta, \vk) +
\sigma_\nu(\eta, \vk) \; , \label{Iecpsi}
\eea
where $ \rho_\gamma(\eta) $ is the photon energy density, $ \Theta_0 (\eta, \vk) $ the 
photon temperature fluctuations integrated over the $ \vq $ directions,
$ \sigma(\eta, \vk) $ is the anisotropic stress perturbation and
\bea\label{Idefd}
&& \Delta_{dm}(\eta,\vk) \equiv \int \frac{d^3q}{(2 \, \pi)^3} \; E(\eta,q) \;  f_1^{dm}(\eta, \vq, \vk) 
 \quad  ,  \quad  
\Sigma_{dm}(\eta, \vk) =  -2 \int \frac{d^3q}{(2 \, \pi)^3} \; \frac{q^2}{E(\eta,q)}
\; P_2\left( {\check k} \cdot {\cq}\right) \;  f_1^{dm}(\eta, \vq, \vk) \cr \cr 
&& h(\eta)\equiv \frac1{a} \; \frac{da}{d\eta} \quad  ,  \quad
 E(\eta,q) \equiv \sqrt{m^2 \; a^2(\eta) + q^2} \quad  .
\eea
$ P_2(x) $ the Legendre polynomial of order two. Equations analogous to eq.(\ref{Idefd})
hold for neutrinos with the index $ _{dm} $ replaced by $ _\nu $ and $ E(\eta,q) $ 
replaced by $ q $. 

\medskip

The customary DM density contrast $ \delta(\eta,\vk) $ is 
connected with $ \Delta_{dm}(\eta,\vk) $ by
\citep{mab}
\be\label{Ddi}
\delta(\eta,\vk) = \frac{\Delta_{dm}(\eta,\vk)}{\rho_{dm} \; 
[a_{eq} + a(\eta)]} \quad  ,  \quad a_{eq} \simeq \frac1{3200} \; ,
\ee
where $ \rho_{dm} $ is the average DM density today.

\medskip

We start this paper by deriving the collisionless Boltzmann-Vlasov equation
for DM particles which decoupled being ultrarelativistic (UR) and become non-relativistic in the 
radiation dominated era. This treatment is general and applies to any DM particle candidate
decoupling being UR during the RD era. In particular, it is appropriated for keV scale WDM 
particles which become non-relativistic by redshift $ z \sim 5 \times 10^6 $.
Furthermore, we generalize the whole treatment to particles
that decouple being non-relativistic as CDM.

\medskip

Combining the linear and collisionless Boltzmann-Vlasov equations (\ref{bvI})-(\ref{bvnuI}) 
with the linearized Einstein equations (\ref{Iecfidm})-(\ref{Iecpsi}) at initial times
strongly constrain the initial conditions on the distribution functions and the gravitational potentials.
The initial conditions are efficiently investigated expanding the distribution functions
in powers of  $ \eta $ and $ i \; \cq \cdot {\vk} \; \eta $. Our analysis includes the
initial conditions for DM, neutrinos and photons. This analysis is valid both for DM that decouples
being ultrarelativistic and nonrelativistic (as CDM).
We show in this framework that the $ \vk $ dependence  
factorizes out in the initial distribution functions $ \Psi_{dm}(0, \vq, \vk) $ and 
$ \Psi_{\nu}(0, \vq, \vk) $ as well as in the initial densities 
$ \Delta_{dm}(0,\vk), \; \Delta_{\nu}(0,\vk) $ and anisotropic stresses 
$ \sigma_{dm}(0, \vk), \; \sigma_\nu(0, \vk) $. 
The dependence on the directions of $ \vk $ stays factorized
for all times considerably simplifying the dynamical evolution.

\medskip

The primordial inflationary fluctuations \cite{biblia,dod} determine the
initial gravitational potential $ \psi(0, \vk) $.
$ \psi(0, \vk) $ is given by the product of a $ k $ dependent amplitude proportional to $ k^{n_s/2-2} $ 
times a Gaussian random field with unit variance that depends on the $ \vk $-direction, 
$ n_s $ being the scalar primordial index,

\medskip

We derive from the linearized Boltzmann-Vlasov equation (\ref{bvI}) a system of four linear integral
equations of Volterra type for the density fluctuations 
$ \Delta_{dm}(\eta,\vk), \; \Delta_{\nu}(\eta,\vk) $
and the anisotropic stress fluctuations $ \sigma_{dm}(\eta, \vk), \;  \sigma_{\nu}(\eta, \vk) $ 
valid both for
ultrarelativistic and non-relativistic particles in the RD and MD eras. This is a generalization
of Gilbert's equation.  Gilbert's equation is only valid for non-relativistic particles in a matter 
dominated universe \cite{gil}. The remarkable fact in these new Volterra integral equations
is that the density and anisotropic stress fluctuations obey a closed system of integral equations.
Although the B-V equation is an equation on functions of $ \eta, \; \vk $ {\bf and} $ \vq $
with coefficients depending on $ \eta, \; \vk $ {\bf and} $ \vq $, 
integrating the distribution functions on $ \vq $ with appropriated weights,
the density and anisotropic stress fluctuations obey a {\bf closed} system of integral equations. 
Namely, no extra information on the $ \vq $ dependence of the distribution functions is needed,
which is a {\bf truly remarkable} fact.

\medskip

In summary, the pair of partial differential Boltzmann-Vlasov
equations in seven variables eqs.(\ref{bvI}) and (\ref{bvnuI}) becomes a system
of four Volterra linear integral equations on $ \Delta_{dm}(\eta,\vk), \; \Sigma_{dm}(\eta,\vk) 
, \; \Delta_{\nu}(\eta,\vk) $ and $ \Sigma_{\nu}(\eta,\vk) $.  
In addition, because we deal with
linear fluctuations evolving on an homogeneous and isotropic cosmology, the 
Volterra kernel turns to be isotropic, independent of the  $ \vk $ directions.
As stated above, the $ \check{k} $ dependence factorizes out and we arrive to a final system of 
{\bf two} Volterra integral equations in two variables:  the modulus $ k $ and the time that we choose
to be the scale factor.

We have thus considerably simplified the original problem: we reduce
a pair of partial differential B-V equations on seven 
variables $ \eta, \; \vq, \; \vx $
into a pair of Volterra integral equations on two variables: $ \eta, \; k $.

\medskip

It is convenient to define dimensionless variables as 
$$
\alpha \equiv \frac{k \; l_{fs}}{\sqrt{I_4^{dm}}} \quad , \quad 
 l_{fs} = \frac2{H_0} \; \frac{T_d}{m} \; \sqrt{\frac{I^{dm}_4}{a_{eq} \; \Omega_{dm}}} \; ,
$$
where $ l_{fs} $ stands for the free-streaming length \cite{kt,bdvs,dvs},
$ T_d $ is the comoving DM decoupling temperature and $ I_4^{dm} $ is the 
dimensionless square velocity dispersion given by
\be\label{dfIn}
I_n^{dm}=\int_0^\infty Q^n \; f_0^{dm}(Q) \; dQ \quad ,  \quad {\rm while} \;  f_0^{dm}(Q) \; 
{\rm is ~normalized} \; 
 {\rm by} \quad I_2^{dm} = 1  \; .
\ee
$ Q $ is the dimensionless momentum $ Q \equiv q/T_d $ whose typical values are of order one.
We choose as time variable
$$
y \equiv  a(\eta)/a_{eq} \simeq 3200 \; a(\eta) \; .
$$
A relevant dimensionless rate emerges: the ratio between the DM particle mass $ m $ 
and the decoupling temperature at equilibration,
$$
\xi_{dm} \equiv \frac{m \; a_{eq}}{T_d} = 4900 \; \frac{m}{\rm keV} \; 
\left(\frac{g_d}{100}\right)^\frac13 \; ,
$$
$ g_d $ being the effective number of UR degrees of freedom at the DM decoupling.
Therefore, $ \xi_{dm} $ is a large number provided the DM is non-relativistic at equilibration.
For $ m $ in the keV scale we have $ \xi_{dm} \sim 5000 $. 

\medskip

DM particles and the lightest neutrino become non-relativistic by a redshift 
\be 
z_{trans} + 1 \equiv \frac{m}{T_d} \simeq 1.57 \times 10^7 \;   
\frac{m}{\rm keV} \; \left(\frac{g_d}{100}\right)^\frac13 \quad 
{\rm for ~ DM ~ particles}\quad ,
\quad z^{\nu}_{trans} = 34 \; \frac{m_{\nu}}{0.05 \; {\rm eV}}  \quad
{\rm for ~ the ~ lightest ~ neutrino} \; .
\ee
$ z_{trans} $ denoting the transition redshift from ultrarelativistic regime to the 
nonrelativistic regime of the DM particles.

\medskip

Since WDM decouples being ultrarelativistic it contributes to radiation for large redshifts
$ z > z_{trans} $. However, WDM turns to produce a small relative correction of the order
$ 1/\xi_{dm} $ to the photons + neutrino density. We find a little slow down of the order 
$ 1/\xi_{dm} $ in the expansion of the universe when the WDM becomes 
non-relativistic around $ \xi_{dm} \; y \simeq 1 $.

\medskip

We obtain a pair of coupled Volterra equations for the functions $ {\breve \Delta}(y,\alpha) $
and $ {\bar \sigma}(y,\alpha) $ defined as follows
\be\label{defV}
{\breve \Delta}(y,\alpha) = -\frac1{2 \, I_\xi} \; \left[\frac1{\xi_{dm}} \;  
{\bar \Delta}_{dm}(y,\alpha) 
+ \frac{R_{\nu}(y)}{I_3^\nu} \;  {\bar \Delta}_{\nu}(y,\alpha) \right] \quad , \quad 
{\bar \sigma}(y,\alpha) = {\bar \phi}(y, \alpha) - 1 \quad ,\quad 
{\bar \phi}(y, \alpha) = \frac{\phi(\eta, \vk)}{\psi(0, \vk)} \quad , \quad 
{\breve \Delta}(0,\alpha) = 1 \; ,
\ee
where we choose to factor out the initial gravitational potential $ \psi(0, \vk) $,
$ R_{\nu}(y) $ is the neutrino fraction of the average energy density,
$$
I_\xi \equiv \displaystyle \frac{I_3^{dm}}{\xi_{dm}} + R_\nu(0)
\quad ,  \quad {\bar \Delta}_{dm}(y,\alpha) = \frac{m}{\rho_{dm} \; T_d \; \psi(0, \vk)} \; 
\Delta_{dm}(\eta,\vk) \quad ,  \quad  {\bar \Delta}_{\nu}(y,\alpha) = 
\frac{I_3^{\nu}}{\rho_r\; \psi(0, \vk) \;  R_{\nu}(y)} \; \Delta_\nu(\eta,\vk) \; ,
$$
$ \Delta_{dm}(\eta,\vk) $ being given by eq.(\ref{Idefd}). Notice that the DM contribution to 
$ {\breve \Delta}(y,\alpha) $ is suppressed by a factor $ 1/\xi_{dm} \simeq 1/5000 $. 
The growth of the DM fluctuations in the MD era largely overcomes this suppression. 

\medskip

Expanding the Boltzmann-Vlasov equations (\ref{bvI}) and (\ref{bvnuI}) in powers of $ \eta $ and 
$ i \; \cq \cdot {\vk} \; \eta $ as remarked above, we obtain the initial gravitational potentials 
related by
$$
\phi(0,\vk) = \left[1 + \frac25 \; I_\xi \right]  \psi(0,\vk) \simeq
\left[1 + \frac25 \; R_\nu(0) \right]  \psi(0,\vk) \; .
$$
Notice above the small $ {\cal O}(1/\xi_{dm}) $ correction in $ I_\xi $.

The final pair of dimensionless Volterra integral equations take the form
\bea\label{Ifinal}
&&{\breve \Delta}(y,\alpha) =  C(y,\alpha) + B_\xi(y)  \; {\bar \phi}(y,\alpha) +
\int_0^y dy' \left[G_\alpha(y,y') \; {\bar \phi}(y',\alpha) +
G^\sigma_\alpha(y,y') \; {\bar \sigma}(y',\alpha)\right] \; , \\ \cr
&& {\bar \sigma}(y,\alpha) =  C^\sigma(y,\alpha) + \int_0^y dy'  \left[
I^\sigma_\alpha(y,y') \;  {\bar \sigma}(y',\alpha) + I_\alpha(y,y') \; {\bar \phi}(y',\alpha)
\right] \; ,\label{Ifinal2}
\eea
with initial conditions $ {\breve \Delta}(0,\alpha) = 1 \quad , \quad {\bar \sigma}(0,\alpha) 
= \frac25 \; I_\xi \quad .$ 
This pair of Volterra equations is coupled with the linearized Einstein equations
(\ref{Iecfidm})-(\ref{Iecpsi}).

\medskip

The kernels and the inhomogeneous terms in eqs.(\ref{Ifinal})-(\ref{Ifinal2})
are given explicitly by eqs.(\ref{varnor})-(\ref{bynsi}), (\ref{asd4})-(\ref{uaysi})
and (\ref{gorda2})-(\ref{defialfa}). The arguments of these functions contain the 
dimensionless free-streaming distance $ l(y,Q) $,
\be\label{lfsI}
l(y,Q) = \int_0^y \frac{dy'}{\sqrt{\left[1+ y'\right] \; \left[ y'^2 + \displaystyle 
\left(\displaystyle Q/\xi_{dm}\right)^2 \right]}}  \; .
\ee
\vskip -0.2 cm
The coupled Volterra integral equations (\ref{Ifinal})-(\ref{Ifinal2}) are easily amenable to a 
numerical treatment. 

\medskip

When the anisotropic stress $ {\bar \sigma}(y, \alpha) $ is negligible,
eqs.(\ref{Ifinal})-(\ref{Ifinal2}) reduce to a single Volterra integral 
equation for the DM density fluctuations $ {\breve \Delta}_{dm}(y,\alpha) $ .
We find the solution of this single Volterra equation for a broad range of 
wavenumbers $ 0.1 / {\rm Mpc} < k < 1/5 \; {\rm kpc} $ and analyze the transfert function and density
contrast in the accompanying paper ref. \cite{dos}.

\medskip

The framework derived here reducing the full evolution of the primordial
cosmological fluctuations to a pair Volterra integral equations is general 
for any type of DM and provides in particular, in the nonrelativistic 
limit in the MD era the so-called Gilbert equations. 

\medskip

In summary, the Volterra integral equations obtained here provide a useful 
and precise method to compute the primordial DM fluctuations (both WDM and 
CDM) over a wide range of scales including very small scales up to 5 kpc.

It is easy to introduce the cosmological constant in the framework and 
equations
presented here. Moreover, baryons and photons can be treated in this framework
at the price of introducing further coupled Volterra integral equations.

\medskip

In section \ref{bolvla} we derive the linearized and collisionless 
Boltzmann-Vlasov (B-V) equations for DM and for neutrinos. 
We present the linearized Einstein equations for the gravitational
potentials which are coupled to the B-V equations. 

In section \ref{condin1} we then provide the adiabatic initial conditions
for the fluctuations which turns to be constrained by the B-V and linearized
Einstein equations.

In section \ref{volt} we recast  the linearized DM and neutrino B-V 
equations as a system of linear integral equations of Volterra type. 

\medskip

We derive in Appendix \ref{ailam} the Poisson equation from the explicit solution of the 
linearized Einstein equations and the systematic corrections to it in the short 
wavelength regime ($ \xi_{dm} \; \alpha  \; y \gg 1 $). Some useful angular integrals are computed in 
Appendix \ref{angus}.

\section{The Boltzmann-Vlasov equation in the FRW universe}\label{bolvla}

We derive in this section the collisionless B-V equation for DM particles which decoupled being 
ultrarelativistic (UR) and become non-relativistic in the 
radiation dominated era. This treatment is appropriate for keV scale DM particles which 
become non-relativistic by $ z \sim 2 \times 10^7 $ and applies also to any DM particle candidate
decoupling being UR during the RD era. 

\subsection{Particle propagation in the FRW universe including
fluctuations}

We consider spatially flat FRW spacetimes with adiabatic
perturbations of the metric in the conformal gauge. In conformal time
$ \eta $ the metric takes the form
\be\label{FRWf}
ds^2= -a^2(\eta) \; \left[1 + 2 \; {\tilde \psi}(\eta,x^i)  \right] \; d\eta^2 +
a^2(\eta) \; \left[1 - 2 \; {\tilde \phi}(\eta,x^i) \right] \; (dx^i)^2 \quad ,
\ee
where $ {\tilde \psi} $ is the Newtonian potential
and $ {\tilde \phi} $ corresponds to the perturbation of the spatial curvature.

\medskip

The particle propagation equations follow from the Lagrangean
\be\label{lagra}
{\cal L} = \frac12 \; g_{\alpha \, \beta} \; \frac{dx^\alpha}{d\lambda}
\; \frac{dx^\beta}{d\lambda} \; ,
\ee
where $ x^\alpha $ are the contravariant particle coordinates and 
$ \lambda $ is the affine parameter on the trajectory.

\medskip

The covariant canonical momentum follows from eq.(\ref{lagra}) as
\be\label{pcov}
p_\alpha \equiv \frac{\partial{\cal L}}{\partial\left(
\displaystyle\frac{dx^\alpha}{d\lambda}\right)} =
g_{\alpha \, \beta} \; \frac{dx^\beta}{d\lambda}=
g_{\alpha \, \beta} \; p^{\beta} 
\ee
and the equations of motion take the form
\be\label{dp}
\frac{dp_\alpha}{d\lambda}= - \frac12 \; p_\beta \; p_\gamma \;
\frac{\partial g^{\beta \, \gamma}}{\partial x^\alpha} \; .
\ee
The equations of motion have to be supplemented by the on-shell condition
\be\label{masa}
m^2 = -g_{\alpha \, \beta} \; p^\alpha \; p^\beta \; ,
\ee
where $ m $ is the mass of the DM particle.

\medskip

The derivative with respect to $ \lambda $ is related to
the derivative with respect to the conformal time using
eq.(\ref{pcov}) for $ \alpha = 0 $:
\be\label{detad}
\frac{d}{d\eta} = \frac1{p^0} \; \frac{d}{d\lambda} \; .
\ee
The equations of motion (\ref{dp}) are then
\be\label{eqpj}
\frac{dp_j}{d\eta} = - \frac1{ a^2(\eta) \; p^0} \; \left(p_i^2 \; \partial_j {\tilde \phi}
+ p_0^2  \; \partial_j {\tilde \psi}\right) \quad .
\ee
To first order in the fluctuations, 
it is convenient to define the momentum $ q_j $ and the energy variable 
$ E(\eta,q) $ as in refs. \cite{primeros,mab},
\be\label{dfqE}
q_j \equiv \left(1 + {\tilde \phi} \right) \; p_j \quad , \quad E(\eta,q) \equiv 
\sqrt{m^2 \; a^2(\eta) + q^2} \quad {\rm where} \quad q_i = q^i \quad , \quad
q^2 \equiv q_j \; q_j \; .
\ee
The on-shell condition eq.(\ref{masa}) becomes to first order in the 
fluctuations,
$$
m^2 \; a^2(\eta) = p_0^2 - p_i^2 - 2 \; \left(p_0^2 \; {\tilde \psi}
+ p_i^2\; {\tilde \phi} \right) 
\quad {\rm or} \quad E^2 =  p_0^2\left(1 - 2 \; {\tilde \psi} \right) \; .
$$
We therefore have to first order in $ {\tilde \phi} $,
\be\label{pqE}
p^j = \frac1{ a^2(\eta)} \; \left(1 + {\tilde \phi} \right) \; q^j
\quad , \quad p^0 = \frac1{ a^2(\eta)} \; \left(1 - {\tilde \phi} \right) \; E(\eta,q)
\; ,
\ee
$$
E(\eta,q) = -p_0 \; \left(1 - {\tilde \phi} \right) \quad , \quad 
p^2 = p_i \; p^i = \frac{q^2}{a^2(\eta)} \quad , \quad p_i^2 + p_0^2 =
E^2(\eta,q) + q^2 + 2 \; {\tilde \phi} \; m^2 \; a^2(\eta) \; .
$$
In terms of $ q_i $ the equations of motion (\ref{eqpj}) 
to first order in the fluctuations $ {\tilde \phi} $ take the form
\be
\frac{d q_i}{d \eta} = q_i \; 
\frac{d {\tilde \phi}}{d \eta} - 
\left(E \; \partial_i {\tilde \psi}+ \frac{q^2}{E} \; \partial_i {\tilde \phi} \right)
\ee
and therefore
\be\label{dqeta1}
\frac{d q}{d\eta} =\frac{q_i}{q} \; \frac{d q_i}{d \eta}=
  q \; \frac{d {\tilde \phi}}{d \eta} - \left(E  \; n_i \; \partial_i {\tilde \psi}+ 
\frac{q^2}{E}  \; n_i \; \partial_i {\tilde \phi} \right) \; ,
\ee
where 
\be\label{dni}
n_i \equiv \frac{q_i}{q} \quad , \quad n_i \; n^i = \delta^{i \, j} \; n_i \; n_j = 1 
 \quad , \quad  n_i \; \frac{d n^i}{d\eta} = 0 \; .
\ee
The total derivative $ d{\tilde \phi}/d\eta $ can be expressed 
in terms of partial derivatives as
\be\label{trampa}
\frac{d {\tilde \phi}}{d \eta} = \frac{\partial {\tilde \phi}}{\partial \eta}
+ \frac{q_i}{E} \; \partial_i {\tilde \phi} 
\ee
where from eqs.(\ref{pcov}), (\ref{detad}) and (\ref{dfqE})
to zeroth order in the fluctuations,
\be\label{xpu}
\frac{d x^i}{d\eta} = \frac{p^i}{p^0} =\frac{q_i}{E} \; .
\ee
Combining eqs.(\ref{dqeta1}) and (\ref{trampa}) yields 
\bea\label{dqetai}
&& \frac{d q_i}{d \eta} = q_i \; \frac{\partial {\tilde \phi}}{\partial \eta}
- E \; \partial_i {\tilde \psi}  - \frac{q^2}{E}\left(\delta_{i \, j}
- n_i \; n_j \right)\partial_j {\tilde \phi} \\ \cr 
&&\frac{d q}{d\eta} = 
 q \; \frac{\partial {\tilde \phi}}{\partial \eta} - E \; n_i \; 
\partial_i {\tilde \psi} \label{dqeta} \; .
\eea

\subsection{The zeroth-order WDM distribution
 and the space-time in the RD and MD eras} 

We work in the universe where radiation and dark matter are both present.
The radiation and DM densities are given in general at zeroth order by
\be\label{rydm}
\rho_r(a) = \frac{\rho_r}{a^4} 
\quad , \quad
\rho_{dm}(a) = \frac{{\hat N}_{dm}}{a^4} \; g_{dm} 
\int \frac{d^3q}{(2 \, \pi)^3} \; E(\eta,q) \; {\hat f^{dm}}_0(q) 
\quad ,
\ee
where $ \rho_r = \Omega_r \; \rho_c $ stands for the radiation energy density today,
$ {\hat f^{dm}}_0(q) $ is the homogeneous and isotropic zeroth order 
distribution that freezed out at decoupling, normalized as
\be\label{nf0}
\int_0^\infty q^2 \; dq \; {\hat f^{dm}}_0(q) = 1 \; .
\ee
$ g_{dm} $ is the number of internal degrees of freedom 
of the DM particle, typically $ 1 \leq g_{dm} \leq 4 $
and the normalization factor $ {\hat N_{dm}}$ reproduces the DM average 
density today $ \rho_{dm} = \Omega_{dm} \; \rho_c $ as
\be\label{Nsom}
{\hat N_{dm}}\; m \; g_{dm} \int \frac{d^3q}{(2\pi)^3} \;  {\hat f^{dm}}_0(q)  = \Omega_{dm} \; \rho_c 
\quad ,  \quad  {\rm hence,} \quad 
{\hat N_{dm}}= \frac{2 \, \pi^2 \; \rho_{dm}}{g_{dm} \; m} \; ,
\ee
where $ \Omega_{dm} =  0.233 $ is the DM fraction and $ \rho_c $ is the 
critical density of the Universe
\be\label{roc}
\rho_c = 3 \, M_{Pl}^2 \; H_0^2 = (2.518 \; {\rm meV})^4 \quad , \quad 
1 \, {\rm meV} = 10^{-3} \, {\rm eV} \quad , \quad H_0 = 1.5028 \; 10^{-42} 
\; {\rm GeV} \quad , \quad  M_{Pl}^2 = \frac1{8 \, \pi \; G} \; .
\ee

We consider {\bf generic} freezed out distribution functions $ {\hat f^{dm}}_0(q) $. 
We call $ T_d $ the scale of the average momentum $ q $ at the zeroth order 
freezed-out homogeneous and isotropic distribution at decoupling. 
When the DM particles decouple at thermal equilibrium, $ T_d $ is just the (covariant)
decoupling temperature. $ T_d $  is related by entropy conservation to the CMB temperature today
and to the effective number of UR degrees of freedom at decoupling $ g_d $ as
\be\label{temp}
T_d = \left(\frac2{g_d}\right)^\frac13 \; T_{cmb}  \;
, \quad {\rm where} \quad T_{cmb} = 0.2348 \; {\rm meV} \; .  
\ee
In case the decoupling happens out of thermal equilibrium, 
$ T_d $ gives the (covariant) momentum scale of the DM particles at decoupling.
We thus introduce the dimensionless momentum both for in and out of equilibrium decoupling,
\be\label{defQ}
Q \equiv \frac{q}{T_d} \; ,
\ee
which typical values are of order one.

We now consider the dimensionless zeroth order freezed-out density 
$ f_0^{dm}(Q) $ and the dimensionless normalization constant $ N_{dm} $
\be\label{defnor}
 f_0^{dm}(Q) = T_d^3 \; {\hat f^{dm}}_0(q) \quad , \quad
\int_0^\infty Q^2 \; dQ \; f_0^{dm}(Q) = 1  \quad , \quad
N_{dm} = \frac{\hat N_{dm}}{T_d^3} = 
\frac{2 \; \pi^2 \; \rho_{dm}}{g_{dm} \; m \; T_d^3} \; ,
\ee
where we used eqs.(\ref{nf0}) and (\ref{Nsom}). For example, we have 
for DM fermions decoupling ultrarelativistically at thermal equilibrium 
\be\label{fdsd}
f_0^{dm}(Q) = \frac2{3 \, \zeta(3)} \; \frac1{e^Q + 1} \; ,
\ee
where $ \zeta(3) = 1.2020569\ldots $. Out of equilibrium freezed-out distribution functions 
for sterile neutrinos \cite{dw,neus,bwu,kus,fuera}
are considered in the accompanying paper \cite{dos}.  

\medskip

Eq.(\ref{defnor}) and the value of  the average DM density $ \rho_{dm} $ 
eq.(\ref{roc}) imposes on the parameters of the DM particle:
$$
g_{dm} \;  N_{dm} \; m = 0.6988 \; {\rm keV} \;  \frac{g_d}{100} \; .
$$
This relation suggests that DM decoupling ultrarelativistically can have 
its mass in the keV scale. 
Moreover, an increasing body of evidence from the combination of theory
and astronomical observations points towards DM particles
with mass in the keV scale \cite{bdvs2,dvs,dvs2}: we thus take 1 keV 
as the reference scale for the mass of DM particles. 
We consider $ g_d = 100 $ as reference value for the number $ g_d $ of
ultrarelativistic degrees of freedom at decoupling in thermal equilibrium. This
corresponds to a physical decoupling temperature 
$ T_{d \, phys} = (z_d + 1) \; T_d \sim 100 $ 
GeV, $ T_d $ being the covariant decoupling temperature.

\medskip

The normalized momenta $ I_n^{dm} $ for fermions in thermal equilibrium and 
for out of equilibrium sterile neutrinos are defined as
\be\label{defin}
I_n^{dm} \equiv \int_0^\infty Q^n \; f_0^{dm}(Q) \; dQ \quad , \quad 
I^\nu_n \equiv \int_0^\infty Q^n \; f_0^\nu(Q) \; dQ \; .
\ee
Explicit expressions for them are given in the accompanying paper \cite{dos}.

\begin{table}
\begin{tabular}{|c|} \hline  
Some useful formulas
  \\
\hline \hline
 \\
$ H_0^2 = \displaystyle \frac{8 \; \pi \; G}3 \; \rho_c \quad , \quad M_{Pl}^2 
= \displaystyle \frac1{8 \; \pi \; G} \quad , \quad 
\rho_{dm} =  \Omega_{dm} \; \rho_c  \quad , \quad \rho_r =  \Omega_r \; \rho_c
\quad , \quad
\displaystyle \frac1{a_{eq}} = \displaystyle \frac{\Omega_M}{\Omega_r} \simeq 3200$  
\\  \\ \hline  \\
$ \eta^* = \displaystyle \sqrt{\frac{a_{eq}}{\Omega_M}} \; 
\displaystyle \frac1{H_0} \simeq 143 $ Mpc $ \quad , \quad \kappa = k \; \eta^* = 
\displaystyle \frac12 \; \xi_{dm} \; \alpha \quad , \quad
 g_{dm} \; N_{dm} = 2 \; \pi^2 \; \displaystyle \frac{\rho_{dm}}{m \; T_d^3}\quad , \quad
y = \frac{a}{a_{eq}} \simeq \frac{3200}{z+1}$
 \\ \\ \hline   \\ 
$\xi_{dm} = \displaystyle \frac{m \; a_{eq}}{T_d}= 4900 \; \displaystyle \frac{m}{\rm keV} \; 
\left(\displaystyle \frac{g_d}{100}\right)^\frac13 = 5520 \; \left(\frac{m}{\rm keV}\right)^\frac43 \; 
(g_{dm} \; N_{dm})^\frac13 \quad , \quad
\displaystyle\frac{4 \, \pi \; G \; {\eta^*}^2}{3 \, a_{eq}^2} = \frac1{2 \; \rho_r}$ 
\\  \\ \hline \\
$\alpha =  \displaystyle \frac1{\displaystyle \sqrt{a_{eq} \; \Omega_{dm}}} \; 
\displaystyle\frac{2 \; T_d}{m \; H_0} \; k \quad , \quad
\displaystyle\frac{4 \, \pi \; G \; {\eta^*}^2}{3 \, a_{eq}^2} \; \displaystyle
\frac{g_{dm} \; N_{dm} \; T_d^4}{2 \, \pi^2} = \displaystyle \frac1{2 \; \xi_{dm}}
\quad , \quad \displaystyle\frac{4 \, \pi \; G \; {\eta^*}^2}{3 \, a_{eq}^2} \; \displaystyle
\frac{g_{\nu} \; N_{\nu}(y) \; (T_d^{\nu})^4}{2 \, \pi^2} =  
\displaystyle\frac{R_{\nu}(y)}{2 \, I_3^{\nu}}$
\\  \\ \hline \\
$\varepsilon(y,Q) = \sqrt{(\xi_{dm})^2 \; y^2 + Q^2} \quad , \quad
\varepsilon_{\nu}(y,Q) = Q$
for $ z > 95 \; \displaystyle \frac{m_{\nu}}{0.05 \; {\rm eV}}\quad , \quad
\beta_{\kappa} (y,y') = \left( \displaystyle \frac{1+y}{1+y'} \; 
e^{y'-y} \right)^{\displaystyle \kappa^2 /3}$
\\  \\ \hline \\
$l(y,Q) = \displaystyle \int_0^{y} \frac{dy'}{\displaystyle \sqrt{(1+y') \; 
\left[\displaystyle y'^2 +  \left(\displaystyle Q/\xi_{dm}\right)^2 \right]}} 
\quad , \quad l_Q(y,y') \equiv \frac12 \; Q \left[ l(y,Q)-l(y',Q) \right]$
\\  \\ \hline \\
$\displaystyle I_n^{dm} = \int_0^\infty Q^n \; f_0^{dm}(Q) \; dQ \quad , \quad 
\displaystyle I^\nu_n = \int_0^\infty Q^n \; f_0^\nu(Q)\quad , \quad I_2^{dm} = I^\nu_2 = 1$
\\  \\ \hline \\
$\Delta_{dm}(\eta,\vk) = {\bar \Delta}_{dm}(y,\alpha) \; \displaystyle 
\frac{g_{dm} \; N_{dm} \; T_d^4}{2 \, \pi^2} \; \psi(0,\vk) \quad , \quad 
\Delta_\nu(\eta,\vk) = {\bar \Delta}^\nu(y,\alpha) \; \displaystyle 
\frac{g^\nu \; N^\nu(y) \; (T^\nu_d)^4}{2 \, \pi^2} \; \psi(0,\vk)$
\\  \\ 
${\breve \Delta}(y,\alpha) = -\displaystyle\frac1{2 \, I_\xi} 
\left[ \displaystyle\frac1{\xi_{dm}} \; {\bar \Delta}_{dm}(y,\alpha) 
+  \displaystyle\frac{R_\nu(y)}{I_3^\nu} \; {\bar \Delta}_{\nu}(y,\alpha) \right]
\quad , \quad I_\xi = \displaystyle\frac{I_3^{dm}}{\xi_{dm}} + R_\nu(0) 
\simeq R_{\nu}(0) =  0.727 \quad , \quad {\breve \Delta}(0,\alpha) = 1$
\\ \\ \hline \\
$\phi(\eta,\vk)=\psi(0,\vk) \; {\bar \phi}(y,\alpha) \quad , \quad \psi(\eta,\vk)=
\psi(0,\vk) \; {\breve \psi}(y,\alpha)
\quad , \quad {\breve \psi}(0,\alpha) = 1 \quad , \quad {\bar \phi}(0) \simeq 1 + \frac25 \; R_\nu(0) = 1.291$
\\ \\
$ {\bar \sigma}(y,\alpha) = {\bar \sigma}_{dm}(y,\alpha) + {\bar \sigma}_{\nu}(y,\alpha) \quad , \quad
\sigma(\eta,\vk) = \psi(0,\vk) \; {\bar \sigma}(y,\alpha) 
\quad , \quad \sigma_{dm}(\eta,\vk) = \psi(0,\vk) \; {\bar \sigma}_{dm}(y,\alpha) 
\quad , \quad \sigma_{\nu}(\eta,\vk) = \psi(0,\vk) \; {\bar \sigma}_{\nu}(y,\alpha)$
\\ \\ \hline \\
$r(y,y') = 2 \, \left(\sqrt{1+y}-\sqrt{1+y'} \right) \quad , \quad
s(y) = -{\rm Arg \, Sinh}\left( \displaystyle  \frac1{\sqrt{y}} \right)$
\\ \\ \hline \\
$b_{dm}(y) \buildrel{y \to 0}\over= \displaystyle\frac{4 \; I_3^{dm}}{\xi_{dm} \; y} + 
I_1^{dm} \; \xi_{dm} \; y + {\cal O}(y^3) \quad , \quad
b_{dm}(y)\buildrel{y \gg 1}\over= 3 + \frac{5 \, I_4^{dm}}{2 \, [\xi_{dm} \; y]^2}+
{\cal O}\left(\displaystyle\frac1{[\xi_{dm} \; y]^4}\right)$ 
\\  \\ \hline
\end{tabular}
\caption{Some useful formulas.}
\label{formu}
\end{table}

\begin{table}
\begin{tabular}{|c|c|c|} \hline  
 & & \\
Universe Event & redshift $ z $ & $ y = \displaystyle \frac{a}{a_{eq}}=\displaystyle
\frac{z_{eq}+1}{z+1} \simeq \frac{3200}{z+1}$ \\ 
 & & \\
\hline \hline
 & & \\
DM decoupling & $z_d \sim 1.6 \; 10^{15} \; \frac{T_{dp}}{100 \; {\rm GeV}} \; 
\left(\frac{g_d}{100}\right)^\frac13$ & $ y_d \simeq 2 \times 10^{-12}$  \\ 
 & & \\ \hline 
 & & \\
 neutrino decoupling & $z^{\nu}_d \simeq 6 \times 10^9$ & $ y^{\nu}_d \simeq 0.5 \times 10^{-6} $ \\
 & & \\ \hline 
 & & \\
DM particles transition from UR to NR & $z_{trans} \simeq 1.6 \times 10^7 \; \frac{\rm keV}{m} \;
\left(\frac{g_d}{100}\right)^\frac13$ & $y_{trans} = \displaystyle \frac1{\xi_{dm}} \simeq 0.0002$ \\
 & &  $ 10^{-6} < y < 0.01 $ \\ \hline
 & & \\
Transition from the RD to the MD era & $ z_{eq} \simeq 3200 $ & $ y_{eq} = 1 $ \\
 & & \\ \hline   
 & & \\
The lightest neutrino becomes NR & $z^{\nu}_{trans} = 95 \; \displaystyle \frac{m_{\nu}}{0.05 \; {\rm eV}}$ & 
$y^{\nu}_{trans} = 34 \; \displaystyle \frac{0.05 \; {\rm eV}}{m_{\nu}}$ \\
 & & \\ \hline   
 & & \\
Today & $ z_0 = 0 $ & $ y_0 \simeq 3200 $  \\
 & & \\ \hline  
\end{tabular}
\caption{Main events in the DM, neutrinos and universe evolution.}
\label{esca}
\end{table}

\medskip

From now on we use for the dimensionless one-particle energy [see (\ref{dfqE})],
\be\label{deftzi}
\varepsilon(y,Q) \equiv  \frac{E(\eta,q)}{T_d} = \sqrt{(\xi_{dm})^2 \; y^2 + Q^2} 
\quad {\rm where}\quad  a = a_{eq} \; y \quad {\rm and} \quad
\xi_{dm} \equiv \frac{m \; a_{eq}}{T_d} \; .
\ee
We find from eqs.(\ref{temp}), (\ref{defnor}) and (\ref{deftzi}),
\be\label{forxi}
\xi_{dm} = \frac{m \; a_{eq}}{T_d} = 4900 \; \frac{m}{\rm keV} \; 
\left(\frac{g_d}{100}\right)^\frac13 = 5520 \; \left(\frac{m}{\rm keV}\right)^\frac43 \; 
(g_{dm} \; N_{dm})^\frac13 \; .
\ee
That is, $ \xi_{dm} $ will normally be a large number $ \xi_{dm} \sim 5000 $.
The parameter $ \xi_{dm} $ is the ratio between the DM particle mass $ m $ and the
physical decoupling temperature at  equilibration redshift $ z_{eq} + 1 = 1/a_{eq} \simeq 3200 $.
Therefore, $ \xi_{dm} $ is a large number provided the DM is non-relativistic at equilibration.

\medskip

\begin{figure}[h]
\begin{center}
\begin{turn}{-90}
\psfrag{"LR.dat"}{$ \log_{10} {\cal R}_0(y) $ vs. $ \log_{10} y $
for a Fermi-Dirac distribution }
\psfrag{"LestR.dat"}{$ \log_{10} {\cal R}_0(y) $ vs. $ \log_{10} y $
for  sterile neutrinos in the $ \chi $ model}
\includegraphics[height=13.cm,width=8.cm]{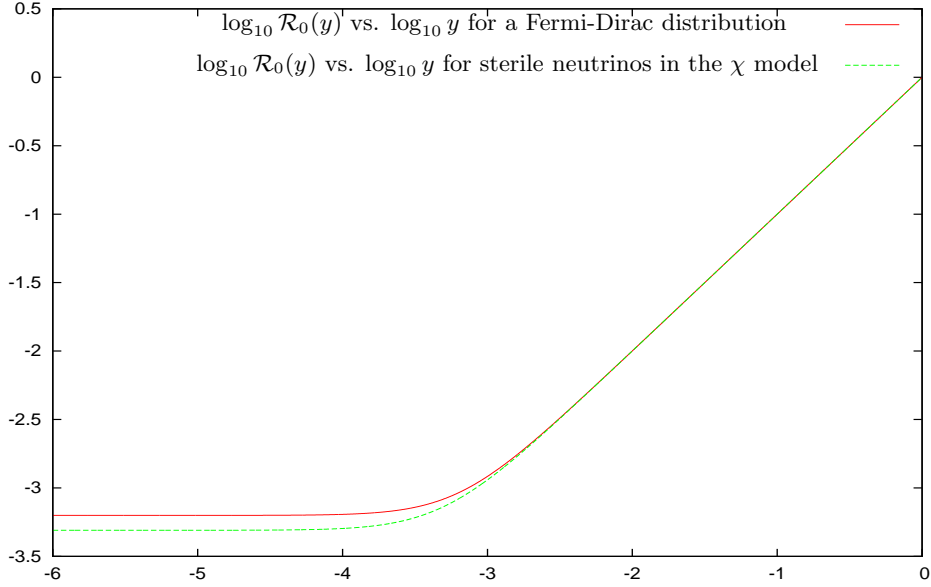}
\end{turn}
\caption{$ \log_{10} {\cal R}_0(y) $ defined in eq.(\ref{defR}) 
vs. $ \log_{10} y $ for fermions in thermal equilibrium and for sterile 
neutrinos out of thermal equilibrium in the $ \chi $ model. Both
freeze-out distributions give the same $ {\cal R}_0(y) $ values for
$ \xi_{dm} \; y \gtrsim 5 $ [as in eq.(\ref{R})] 
while $ {\cal R}_0(y) $ does depend on the
details of the freeze-out distribution for $ \xi_{dm} \; y \lesssim 5 $.
For $ \xi_{dm} \; y \lesssim 1 , \; {\cal R}_0(y) $ takes the constant
value given analytically in eq.(\ref{R}).} 
\label{Rcal}
\end{center}
\end{figure}

It is convenient to use the dimensionless wavenumbers \cite{bdvs}
\be\label{defxi}
\kappa \equiv k \; \eta^* \quad {\rm and} \quad
\alpha \equiv \frac2{\xi_{dm}} \; \kappa = 
\frac2{H_0} \; \frac{T_d}{m \; \sqrt{a_{eq} \; \Omega_{dm}}} \; k \; 
\quad {\rm where} \quad \eta^* \equiv \sqrt{\frac{a_{eq}}{\Omega_M}} \; 
\frac1{H_0} = 143 \; {\rm Mpc} \; .
\ee
The free-streaming length is given by \cite{bdvs,dvs} 
\be\label{deffs}
l_{fs} = \frac2{H_0} \; \frac{T_d}{m} \; \sqrt{\frac{I_4^{dm}}{a_{eq} \; 
\Omega_{dm}}} =  \frac{2 \, \eta^*}{\xi_{dm}} \; \sqrt{I_4^{dm}} \; ,
\ee
where the momenta $ I_n^{dm} $ are defined by eq.(\ref{defin}) and therefore,
\be\label{alfs}
\alpha = \frac{k \; l_{fs}}{\sqrt{I_4^{dm}}} \quad {\rm and} \quad l_{fs} = 57.2 \, {\rm kpc}
\; \frac{\rm keV}{m} \; \left(\frac{100}{g_d}\right)^\frac13 = 
50.8 \, {\rm kpc}
\left(\frac{\rm keV}{m}\right)^\frac43 \; (g_{dm} \; N_{dm})^{-\frac13} \; .
\ee
The DM energy density is given in general by eq.(\ref{rydm})
that we can write as
\be\label{denwdm}
\rho_{dm}(y) = \frac{\rho_{dm}}{a^3(y)} \; \frac{{\cal R}_0(y)}{y} \quad ,
\ee
$ y $ is defined in eq.(\ref{deftzi}), 
$ a_{eq} =  \Omega_r / \Omega_M \simeq 1/3200 $ is the scale factor at 
equilibration, 
\bea\label{defR}
&& {\cal R}_0(y) \equiv \frac{\rho_{dm}(y)}{\rho_r(y)} = 
\int_0^{\infty} Q^2 \; dQ \; \sqrt{y^2 + \frac{Q^2}{\xi_{dm}^2}} \;
f_0^{dm}(Q) \quad , \quad 
\rho_r(y) = \frac{\rho_r}{a^4(y)} \quad  {\rm and}\\ \cr\cr
&& {\cal R}_0(y) =\left\{\begin{array}{l} \displaystyle
\frac{I_3^{dm}}{\xi_{dm}}
\left[1 + {\cal O}\left(\xi_{dm}^2 \; y^2\right) \right]
\quad , \quad \xi_{dm} \; y \lesssim 1 \; , \\
y + \displaystyle \frac{I_4^{dm}}{2 \, \xi_{dm}^2 \; y} +  
{\cal O}\left(\frac1{\xi_{dm}^4 \; y^3}\right)\quad , 
\quad \xi_{dm} \; y \gtrsim 5 \; .
\end{array} \right.\label{R}
\eea
When $ \xi_{dm} \; y \gtrsim 1 $ and the WDM particles are nonrelativistic
the WDM density from eqs.(\ref{denwdm}) and (\ref{R}) dilutes as $ 1/a^3 $
as expected. For $ \xi_{dm} \; y \lesssim 1 $ the WDM particles are 
ultra-relativistic and from eqs.(\ref{denwdm}) and (\ref{R}) 
the WDM density dilutes as radiation as $ 1/a^4 $.
Eq.(\ref{R}) shows that $ \rho_{dm}(y) $ and $ \rho_{rad}(y) $ become 
equal at equilibration $ y = 1 $ (up to $ 1/\xi_{dm}^2 $ corrections), 
as it must be. In fig. \ref{Rcal} we plot $ \log_{10} {\cal R}_0(y) $ vs. 
$ \log_{10} y $ for fermions decoupling in thermal equilibrium
and for sterile neutrinos decoupling out of thermal equilibrium in 
the $ \chi $ model where sterile neutrinos are produced by the decay of a 
real scalar \cite{modelos,kus}. 
(These particle models are analogous to those 
in refs. \cite{complejo} which consider a complex scalar field.)

\medskip

We find from eqs.(\ref{denwdm}), (\ref{defR}) and (\ref{R}) that WDM 
gives a small contribution of the order $ 1/\xi_{dm} $
to the radiation density for $ \xi_{dm} \; y \lesssim 1 $:
$$
\rho_{dm}(y) = \rho_r(y) \; {\cal R}_0(y) 
\buildrel{\xi_{dm} \; y \lesssim 1}\over= 
\frac{I_3^{dm}}{\xi_{dm}} \; \rho_r(y) \; .
$$
That is, the quantity $ \xi_{dm} $ gives the order of magnitude of the
ratio of densities $ \rho_r(y)/\rho_{dm}(y) $ for $ \xi_{dm} \; y \lesssim 1 $
while the WDM is still relativistic.

\medskip

Taking into account eq.(\ref{denwdm}) the Friedmann equation 
takes the form
\be\label{fr2}
a_{eq}^2 \; \left(\frac{dy}{d\eta}\right)^2 = H_0^2 \; \Omega_r \; 
\left[1 + {\cal R}_0(y) \right] \; ,
\ee
with the explicit solution
\be\label{etaexa}
\eta = \eta^* \; \int_0^y \frac{dy'}{\sqrt{1 + {\cal R}_0(y)}} \quad .
\ee
When the WDM particles are UR, we find that they give 
small corrections of the order $ 1/\xi_{dm} $ to
the scale factor $ a(\eta) $
\be\label{awdm}
a(\eta) \buildrel{\xi_{dm} \; y \lesssim 1}\over= 
\frac{a_{eq}}{\eta^*} \sqrt{1 +\frac{I_3^{dm}}{\xi_{dm}}} \; \eta 
\quad , \quad a(\eta) \buildrel{\xi_{dm} \; y \gtrsim 1 \; , \; y\ll 1}\over= a_{eq}
 \; \frac{\eta}{\eta^*} \; .
\ee
Hence, eq.(\ref{awdm}) indicates a little slow down of the order 
$ 1/\xi_{dm} $ in the expansion of the universe when the WDM becomes 
non-relativistic around $ \xi_{dm} \; y \simeq 1 $.
When the WDM particles are NR ($ \xi_{dm} \; y \gtrsim 1 $)
the WDM corrections are even smaller, of the order  $ 1/\xi_{dm}^2 $.

\medskip

In summary, up to small $ 1/\xi_{dm} $ or $ 1/\xi_{dm}^2 $ corrections
for $ \xi_{dm} \; y \lesssim 1 $ or $ \xi_{dm} \; y \gtrsim 1 $,
respectively, the scale factor thus results from eq.(\ref{etaexa}),
\be\label{ay}
a(\eta) = a_{eq} \; y(\eta) \quad , \quad 
y(\eta) = \frac{\eta}{\eta^*}\left(1 + \frac{\eta}{4 \, \eta^*}\right)
\quad , \quad \eta= 2 \, \eta^* \left(\sqrt{1+y}-1\right) \; .
\ee
The scale factor eqs.(\ref{ay}) has the radiation dominated behavior
for $ \eta \ll \eta^* $ and the matter dominated behavior for 
$ \eta \gg \eta^* $. 
Notice that $ y_{eq} = 1  \quad {\rm and} \quad y_{today} \simeq 3200 $.

\medskip

We have for the ratio $ {\cal R}_0(y) $
\be\label{domy}
1+{\cal R}_0(y) =
\left\{\begin{array}{l} \displaystyle
1 + {\cal O}\left(\frac1{\xi_{dm}}\right)
\quad , \quad \xi_{dm} \; y \lesssim 1 \; , \\ \\
 1+y + \displaystyle {\cal O}\left(\frac1{\xi_{dm}^2}\right) 
\quad , \quad \xi_{dm} \; y \gtrsim 5 \; .
\end{array} \right.\; .
\ee
Therefore, we can always approximate $ 1+{\cal R}_0(y) $ by $ 1 + y $ 
because in the case $ \xi_{dm} \; y \lesssim 1, \;  {\cal R}_0(y) \ll 1 $. 
We will therefore replace $ 1+{\cal R}_0(y) $ by $ 1 + y $ in most cases.

\medskip

We obtain for $ h(\eta) $ defined as
\be\label{heta}
h(\eta)\equiv \frac1{a} \; \frac{da}{d\eta} = 
\frac{\sqrt{1+{\cal R}_0(y)}}{\eta^* \; y} = \frac{\sqrt{1+y}}{\eta^* \; y}
\left[ 1 + {\cal O}\left(\frac1{\xi_{dm}}\right)\right] \; ,
\ee
where we used eq.(\ref{etaexa}) and (\ref{domy}).

\medskip

Modes reenter the horizon when their physical wavenumber $ k_{reenter}/a $ 
is equal to the inverse of the Hubble radius $ H = h / a $, that is,
\be\label{kree}
k_{reenter} =  \frac{\sqrt{1+y}}{\eta^* \; y} = 
\frac{\sqrt{1+y}}{y} \; \frac1{1.43 \; 10^5 \; {\rm kpc}} \; .
\ee

\subsection{The linear and collisionless Boltzmann-Vlasov equation for DM
and neutrinos}\label{2b}

The distribution function $ {\tilde f}_{dm}(\eta, \vq, \vx ) $ of the DM particles after their
decoupling is described by the collisionless B-V equation.
The distribution function is thus a constant over the particle 
trajectories (Liouville):
\be\label{bv1}
0 = \frac{d{\tilde f}_{dm}}{d\eta} = \frac{\partial {\tilde f}_{dm}}{\partial \eta} + 
\frac{d q_i}{d\eta} \; \frac{\partial {\tilde f}_{dm}}{\partial q_i} + 
\frac{dx^i}{d\eta} \;\frac{\partial {\tilde f}_{dm}}{\partial x^i} \; ,
\ee
$ \eta, \; q_i = \vq_i , \; \vx^i = x^i $ being the independent variables 
in the distribution function.

\medskip

To linear order in the fluctuations the distribution function of the 
decoupled particles can be written as
\be\label{bollin}
{\tilde f}_{dm}(\eta, \vq, \vx) = {\hat N_{dm}}\; g_{dm} \; {\hat f^{dm}}_0(q) + 
{\tilde f^{dm}}_1(\eta, \vq, \vx) = 
{\hat N_{dm}}\; {\hat f^{dm}}_0(q) \; g_{dm} \left[ 1 + {\tilde \Psi}_{dm}(\eta, \vq, \vx) \right] \; ,
\ee
Terms of order higher than one in $ {\tilde f^{dm}}_1 $ are neglected in the linear 
B-V equation. We have from eq.(\ref{bollin})
\be\label{f1Psi}
{\tilde f}_1^{dm}(\eta, \vq, \vx) = {\hat N_{dm}}\; g_{dm} \; {\hat f^{dm}}_0(q) \; 
{\tilde \Psi}_{dm}(\eta, \vq, \vx) \; .
\ee
Since $ dq_i/d\eta, \; \partial {\tilde f}_{dm}/\partial x^i $ and $ \partial {\tilde f}_{dm}/\partial n_i $
are of order one [see eqs.(\ref{dqetai}) and (\ref{bollin})], we can write eq.(\ref{bv1}) 
to the first order as
\be\label{bv2}
\frac{\partial {\tilde f}_{dm}}{\partial \eta} + 
\frac{d q}{d\eta} \; \frac{\partial {\tilde f}_{dm}}{\partial q} + 
\frac{q_i}{E} \;\frac{\partial {\tilde f}_{dm}}{\partial x^i} = 0 \; ,
\ee
where we used eq.(\ref{xpu}). Inserting the linearized distribution function eq.(\ref{bollin}) into
eq.(\ref{bv2}) yields,
\be\label{bv3}
\frac{\partial{\tilde \Psi}_{dm}}{\partial \eta} + \frac{q}{E} \;
n_i \; \partial_i {\tilde \Psi}_{dm} + \frac{d\ln {\hat f^{dm}}_0}{d\ln q}
\left[\frac{\partial {\tilde \phi}}{\partial \eta} - \frac{E}{q} \; 
n_i \; \partial_i \; {\tilde \psi} \right] = 0 \; ,
\ee
where we used eq. (\ref{dqeta}). Fourier transforming,
\bea\label{trafou}
&& {\tilde \Psi}_{dm}(\eta, \vq, \vx) = \int \frac{d^3k}{(2 \, \pi)^3} \; e^{i \, \vk
\cdot \vx} \; \Psi_{dm}(\eta, \vq, \vk) \quad , \quad 
{\tilde f^{dm}}_1(\eta, \vq, \vx) = \int \frac{d^3k}{(2 \, \pi)^3} \; e^{i \, \vk
\cdot \vx} \; f_1^{dm}(\eta, \vq, \vk) \; , \\ \cr \cr 
&& {\tilde \phi}(\eta, \vx) = \int \frac{d^3k}{(2 \, \pi)^3} \; e^{i \, \vk
\cdot \vx} \; \phi(\eta, \vk) \qquad , \qquad
 {\tilde \psi}(\eta, \vx) = \int \frac{d^3k}{(2 \, \pi)^3} \; e^{i \, \vk
\cdot \vx} \; \psi(\eta, \vk)  \; , \cr \cr \cr 
&& f_1^{dm}(\eta, \vq, \vk) = {\hat N_{dm}}\; g_{dm} \; {\hat f^{dm}}_0(q) \; 
\Psi_{dm}(\eta, \vq, \vk) \; , \label{trafou2}
\eea
equation (\ref{bv3}) becomes \citep{mab}
\be\label{bv4}
\frac{\partial{\Psi_{dm}}}{\partial \eta} + \frac{i \; q}{E} \;
n_i \; k^i \; \Psi_{dm} + \frac{d\ln {\hat f^{dm}}_0}{d\ln q}
\left[\frac{\partial \phi}{\partial \eta} - \frac{i \; E}{q} \; 
n_i \; k^i \; \psi \right] = 0  \; ,
\ee
or, equivalently
\be\label{bvy}
\frac{\partial{\Psi_{dm}}}{\partial \eta} + \frac{i \; \vq \cdot {\vk}}{E(\eta,q)} \;
\; \Psi_{dm} + \frac{d\ln {\hat f^{dm}}_0}{d\ln q} \left[\frac{\partial \phi}{\partial \eta} - 
i \; E(\eta,q) \; \frac{\vq \cdot {\vk}}{q^2} \; \; \psi \right] = 0  \; .
\ee


Neutrinos are described by a distribution function $ f_{\nu}(\eta, \vq, \vx ) $
obeying after decoupling a Boltzmann-Vlasov equation similar to eq.(\ref{bv1}).
Neutrinos decouple in thermal equilibrium \cite{dod,kt}
at redshift $ z^{\nu}_d \simeq 6 \times 10^9 $.
$ T^{\nu}_d $ is the comoving decoupling temperature of the neutrinos
$ T^{\nu}_d \simeq (1 \, {\rm MeV}/z^{\nu}_d) \simeq 0.17 \; 10^{-3} $ eV.

\medskip

We can linearize the B-V equation around the equilibrium
zeroth order neutrino distribution as
\be\label{bolinu}
{\tilde f}_{\nu}(\eta, \vq, \vx) = {\hat N_{\nu}}(\eta) \; g_{\nu} \; {\hat f}^{\nu}_0(q) 
+ {\tilde f}_1^{\nu}(\eta, \vq, \vx) = 
{\hat N_{\nu}}(\eta) \; {\hat f}^{\nu}_0(q) \; g_{\nu} \left[ 1 + 
{\tilde \Psi}_{\nu}(\eta, \vq, \vx) \right] \; ,
\ee
where 
\be\label{fnu}
{\hat f}^{\nu}_0(q) = \frac2{3 \, \zeta(3) \; (T^{\nu}_d)^3} \; \frac1{e^{q/T^{\nu}_d}+1} \; .
\ee
The normalization of the neutrino distribution eq.(\ref{bolinu}) is fixed by 
the neutrino energy density being a fraction $ R_\nu(\eta) $ of the radiation energy density
$ \rho_r = \Omega_r \; \rho_c $ in the radiation dominated era
\be\label{rnu}
R_\nu(\eta) \; \rho_r = 
{\hat N}_{\nu}(\eta) \; g_{\nu} \; \int \frac{d^3q}{(2 \, \pi)^3} \; q \; {\hat f}^{\nu}_0(q)
= \frac{{\hat N}_{\nu}(\eta) \; g_{\nu}}{2 \, \pi^2} \; \int_0^\infty q^3 \; dq \; {\hat f}^{\nu}_0(q)
\ee
which gives using eq.(\ref{fnu})
\be\label{nnu}
\frac{g_{\nu} \; T^{\nu}_d}{2 \, \pi^2} \; {\hat N}_{\nu}(\eta)= 
\frac{\rho_r}{I^\nu_3} \; R_\nu(\eta) \; ,
\ee
where $ I^\nu_3 = 7 \, \zeta(4)/[2 \, \zeta(3)] $ for the Fermi-Dirac distribution.
The neutrino fraction $ R_\nu(\eta) $ changes at the temperature of electron-positron annhiliation 
(see ref. \cite{gor} and the accompanying paper \cite{dos}) and becomes negligible in the matter dominated era.

\medskip

Neutrinos can be considered massless and otherwise can be neglected. Therefore 
$ \Psi _{\nu}(\eta, \vq, \vx) $ obeys the massless version of eq.(\ref{bv4})
\be\label{bvnu}
\frac{\partial{\Psi_{\nu}}}{\partial \eta} + i \; 
n_i \; k^i \; \Psi_{\nu} + \frac{d\ln {\hat f}^{\nu}_0}{d\ln q}
\left[\frac{\partial \phi}{\partial \eta} - i \; n_i \; k^i \; \psi \right] = 0  \; .
\ee

\subsection{The linearized Einstein equations for the gravitational potentials.}

The Einstein equations for the FRW metric plus fluctuations eq.(\ref{FRWf}) 
give for the gravitational potential at linear order \citep{dod,mab}
\bea\label{ecfi}
&&3 \; h(\eta) \; \frac{\partial \phi}{\partial \eta} + k^2 \; \phi(\eta, \vk)+ 3 \; h^2(\eta)
\; \psi(\eta, \vk) = 4 \, \pi \; G \; a^2(\eta) \; \delta T^0_0(\eta, \vk) \; , \\ \cr
&& k^2 \; \left[ \phi(\eta, \vk) - \psi(\eta, \vk)\right]  = 4 \, \pi \; G \;
\frac{\Sigma(\eta, \vk)}{a^2(\eta)} \; , \label{ecfic}
\eea
where $ h(\eta) $ is defined in eq.(\ref{heta}), $ \delta T^0_0 $ contains the contributions 
to the energy density from the photons, neutrinos and DM fluctuations and 
$ \Sigma(\eta, \vk) $ is the anisotropic stress perturbation.

\medskip

During the RD era radiation dominates over matter and therefore
the DM fluctuations are much smaller than the radiation fluctuations.
Thus, the gravitational potential is dominated by the radiation fluctuations (photons and neutrinos).
The photons can be described in the hydrodynamical approximation (their anisotropic
stress is negligible). 

The tight coupling of the photons to the electron/protons in the plasma 
suppresses before recombination all photon multipoles except
$ \Theta_0 $ and $ \Theta_1 $. (The  $ \Theta_l $ stem from the Legendre
polynomial expansion of the photon temperature fluctuations
$ \Theta(\eta, \vq, \vk) $ \cite{dod}). 

$ \Theta_0 $ and $ \Theta_1 $ obey the hydrodynamical equations \cite{dod}
\bea\label{hidro2}
&& \frac{d\Theta_0}{d \eta}  + k \; \Theta_1(\eta,\vk) = \frac{d\phi}{d \eta} 
\quad , \\ \cr 
&& \frac{d\Theta_1}{d \eta} - \frac{k}3 \; \Theta_0(\eta,\vk)= 
\frac{k}3 \;\phi(\eta,\vk) \label{hidro} \quad . 
\eea
This is a good approximation for the purposes of following the DM evolution \cite{dod}.

\medskip

The energy-momentum fluctuations are the sum of the DM, photons and neutrino contributions
$$
\delta T^0_0(\eta, \vk) = -\frac{\Delta_{dm}(\eta,\vk) + \Delta_\nu(\eta,\vk)}{a^4(\eta)} 
- 4 \, R_\gamma(\eta) \; \rho_r(\eta) \; \Theta_0 (\eta, \vk) \; ,
$$
while only DM and neutrinos contribute to the anisotropic stress $ \Sigma(\eta, \vk) $ 
\be
\Sigma(\eta, \vk)= \Sigma_{dm}(\eta, \vk)+\Sigma_\nu(\eta, \vk) \; .
\ee
$ R_\gamma(\eta) $ stands for the photon fraction of the radiation and
$ \rho_r(\eta) $ for the radiation density (neutrinos plus photons).
$ R_\gamma(\eta) $ vanishes in the MD era.

\medskip

The DM contribution to the energy-momentum tensor and to the anisotropic stress take the form
\bea\label{t00}
\Delta_{dm}(\eta,\vk) &\equiv& \int \frac{d^3q}{(2 \, \pi)^3} \; E(\eta,q) \;  f_1^{dm}(\eta, \vq, \vk) 
=  {\hat N_{dm}}\; g_{dm} \int \frac{d^3q}{(2 \, \pi)^3} \; E(\eta,q) \; {\hat f^{dm}}_0(q) \; 
\Psi_{dm}(\eta, \vq, \vk) \; , \\ \cr \cr
\Sigma_{dm}(\eta, \vk) &=&  \int \frac{d^3q}{(2 \, \pi)^3} \; \frac{q^2}{E(\eta,q)}
\; \left[1 - 3 \,\left( {\check k} \cdot {\cq}\right)^2 \right]  \;  f_1^{dm}(\eta, \vq, \vk) = \\ \cr \cr
&=& - 2 \,  {\hat N_{dm}}\; g_{dm} \int \frac{d^3q}{(2 \, \pi)^3} \; \frac{q^2 \; 
P_2\left( {\check k} \cdot {\cq}\right)}{E(\eta,q)} \;   {\hat f^{dm}}_0(q) \; 
\Psi_{dm}(\eta, \vq, \vk) \; ,\label{sigma} 
\eea
where $ P_2(x) = (3 \, x^2 -1)/2 $ is the Legendre polynomial of order two and
$ \Delta_{dm}(\eta,\vk) $ stands for the DM density fluctuations in general
(whatever ultrarelativistic,  non-relativistic or intermediate regimes).

\medskip

Similarly, the neutrino contributions take the form
\bea\label{ecfinu2}
&& \Delta_{\nu}(\eta,\vk) = 
{\hat N}_{\nu}(\eta) \; g_{\nu} \; \int \frac{d^3q}{(2 \, \pi)^3} \; q \; {\hat f}^{\nu}_0(q) \; 
\Psi_{\nu}(\eta, \vq, \vk) \; , \\ \cr 
&&\Sigma_{\nu}(\eta, \vk) = - 2 \, {\hat N}_{\nu}(\eta) \; g_{\nu} 
\int \frac{d^3q}{(2 \, \pi)^3} \; q \;  {\hat f}^{\nu}_0(q) \;
P_2\left( {\check k} \cdot {\cq}\right) \; \Psi_{\nu}(\eta, \vq, \vk) \; . \label{signu}
\eea
The gravitational potentials $ \phi(\eta), \; \psi(\eta) $ thus obey 
\bea\label{ecfidm}
&&3 \; h(\eta)  \; \frac{\partial \phi}{\partial \eta} + k^2 \; \phi(\eta, \vk) + 3 \; 
h^2(\eta) \; \psi(\eta, \vk) = -4 \, \pi \; G \; \left[\frac{
\Delta_{dm}(\eta,\vk) + \Delta_\nu(\eta,\vk)}{a^2(\eta)} + 4 \,  
a^2(\eta) \; \rho_\gamma(\eta) \; \Theta_0 (\eta, \vk) \right]  \; , \\ \cr 
&& \sigma(\eta, \vk) \equiv  \phi(\eta, \vk) - \psi(\eta, \vk)  = \frac{4 \, \pi \; G }{k^2 \; a^2(\eta)}
\; \left[\Sigma_{dm}(\eta, \vk)+\Sigma_\nu(\eta, \vk) \right] \quad {\rm where} \quad 
\rho_\gamma(\eta) =  R_\gamma(\eta) \; \rho_r(\eta) \; ,
\label{ecpsi}
\eea
as follows from eqs.(\ref{ecfi})-(\ref{signu}).

In the radiation/matter domination eras the gravitational potential equation (\ref{ecfidm}) takes 
in the dimensionless variables $ y $ and $ \vka $ the form,
\be\label{ecfiy}
y[1+{\cal R}_0(y)] \; \frac{\partial \phi}{\partial y} +
\frac13 \left(\kappa \; y \right)^2 \; \phi(y,\vka)+ \left[1 + {\cal R}_0(y) \right] \; \psi(y,\vka) = 
-\frac{4 \, \pi \; G \; {\eta^*}^2}{3 \, a_{eq}^2} \; \left[\Delta_{dm}(y,\vka)+\Delta_{\nu}(y,\vka)\right]
- 2 \; R_\gamma(y) \; \Theta_0 (y, \vka) \; ,
\ee
where $ \kappa $ is defined in eq.(\ref{defxi}) and we used 
\be\label{defka}
16 \, \pi \; G \; a^4(\eta)\; \rho_\gamma(\eta) =  2 \,  R_\gamma(\eta) \; \frac{3 \; a_{eq}^2}{{\eta^*}^2} 
 \quad ,  \quad {\cal R}_0(y) = \frac{\rho_{dm}(y)}{\rho_r(y)} \; .
\ee
$ \Delta_{dm}(\eta,\vk) $ is connected to the customary DM density contrast $ \delta(\eta,\vk) $ by
\citep{mab}
\be\label{Dd}
 \delta(\eta,\vk) \equiv \displaystyle
-\frac{\displaystyle\delta T^0_{0 \; dm}(\eta, \vk)}{\displaystyle \frac{\rho_{dm}}{a^3} + 
\frac{\rho_r}{a^4}}= \frac{\Delta_{dm}(y,\vka)}{\rho_{dm} \; a_{eq} \; (y+1)} \; .
\ee

\medskip

In the short wavelength limit $ k^2 \gg h^2 $, eq.(\ref{ecfidm}) becomes
the Poisson equation, as expected
\be\label{poi}
k^2 \; \phi_{dm}(\eta, \vk)  \buildrel{\rm non-relativistic}\over= 
4 \, \pi \; G \; \rho_{dm} \; \frac{a(\eta) + a_{eq}}{a^2(\eta)} \; \delta(\eta,\vk) \; .
\ee
In Appendix \ref{ailam} we provide the explicit integral representation (\ref{ila})
to the solution of the first order differential equation (\ref{ecfiy}).
Then, we derive the asymptotic expansion of eqs.(\ref{ecfiy}) and (\ref{ila}) 
in the $ \kappa \; y \gg 1 $ (short wavelength) regime. We obtain in this way the Poisson 
equation eq.(\ref{poi}) plus the next to leading terms in this regime in eq.(\ref{apr1}).

Notice that the anisotropic stress  $ \sigma(y, \vk) $ vanishes for $ \kappa \; y \gg 1 $. 

\section{Initial Conditions for the linearized Boltzmann-Vlasov and Einstein equations}
\label{condin1}

We investigate in this section the initial conditions for the DM linearized distribution function
$ \Psi(\eta, \vq, \vk) $ solution of eq.(\ref{bvy}) and the gravitational potentials  $ \phi(y,\vk) $ 
and $ \psi(y,\vk) $ which obey the linearized Einstein equations (\ref{ecfidm}). 

\medskip

Strictly speaking we should take the initial conditions when both neutrinos and dark matter
are decoupled, namely, at $ y = 0.5 \; 10^{-6} $ (see Table II) instead of $ y = 0 $.
However, setting the initial conditions at $ y = 0 $ as we do here 
introduces at most an error of the order $ 10^{-6} $, that we can safely 
ignore, because both the distribution function and its adiabatic
fluctuations (including the gravitational potentials) are regular at $ y = 0 $.

Eq.(\ref{ecfidm}) yields in the $ \eta = 0 $ limit
\be\label{eilin}
\psi(0,\vk) = -\frac{4 \, \pi \; G \; {\eta^*}^2}{3 \, a_{eq}^2} \; 
\left[\Delta_{dm}(0,\vk)+\Delta_{\nu}(0,\vk)\right]- 2 \;  R_\gamma(0) \; \Theta_0 (0, \vk) \; .
\ee
In order $ \phi(\eta,\vk) $ and $ \psi(\eta,\vk) $ be regular at 
$ \eta = 0 $, eq.(\ref{ecpsi}) implies that
\be\label{condS}
\Sigma_{dm}(0,\vk) =0 \quad , \quad \Sigma_{\nu}(0,\vk) =0 \quad 
{\rm and}  \quad \frac{\partial{\Sigma}_{dm}}{\partial \eta}(0,\vk) =0 
\quad , \quad \frac{\partial{\Sigma}_{\nu}}{\partial \eta}(0,\vk) =0 \; .
\ee
These two conditions are fullfilled provided
the integrals over the directions $ \vq $ of $ \Psi_{dm}(0, \vq, \vk), \; 
\partial \Psi_{dm}(0, \vq, \vk)/\partial \eta , \; 
\Psi_\nu(0, \vq, \vk) $ and $\partial \Psi_\nu(0, \vq, \vk)/\partial \eta $
times the Legendre polynomial $ P_2\left( {\check k} \cdot {\cq}\right) $
vanish in eqs.(\ref{sigma}) and (\ref{signu}), respectively.

\medskip

In the $ \eta \to 0 $ limit all fluctuation modes become superhorizon
and therefore adiabatic modes must become $\vq$-independent 
except for the proportionality to the zeroth order distributions \cite{gor}.
In any case, $ \Psi_{dm}(0, \vq, \vk) $ and $ \Psi_\nu(0, \vq, \vk) $
must be independent of the direction of $ \vq $:
\be\label{iniPsi}
\Psi_{dm}(0, \vq, \vk) = \Psi_{dm}(0,q,\vk) \quad {\rm and} \quad
\Psi_\nu(0, \vq, \vk)= \Psi_\nu(0,q,\vk) \; .
\ee
The linearized Boltzmann-Vlasov equation eq.(\ref{bvy}) yields to the order $ \eta^0 $:
\be\label{iniPsi2}
i \; \cq \cdot {\vk} \; \left[\Psi_{dm}(0,q,\vk) -
\frac{d\ln {\hat f^{dm}}_0}{d\ln q} \; \psi(0,\vk)\right] + 
\frac{\partial{\Psi}_{dm}}{\partial \eta}(0,\vq,\vk)+\frac{d\ln {\hat f^{dm}}_0}{d\ln q}  \; 
\frac{\partial{\phi}}{\partial \eta}(0,\vk) = 0 \; ,
\ee
and a similar expression for the neutrino distribution function.
The superhorizon arguments above and eqs.(\ref{iniPsi})-(\ref{iniPsi2})
suggest an expansion in powers of $ \eta $ and
$ i \; \cq \cdot {\vk} \; \eta $ for the distribution function:
\be\label{iniPsi3}
\frac{\partial{\Psi}_{dm}}{\partial \eta}(0,\vq,\vk) = E_{dm}(q,\vk) \;
i \; \cq \cdot {\vk} +  F_{dm}(q,\vk)  \quad ,  \quad
\frac{\partial{\Psi}_{\nu}}{\partial \eta}(0,\vq,\vk) = E_{\nu}(q,\vk) \;
i \; \cq \cdot {\vk} +  F_{\nu}(q,\vk) \; .
\ee
Eq.(\ref{iniPsi2}) determines the coefficients $ E_{dm}(q,\vk) $
and $ F_{dm}(q,\vk) $ as
\be
E_{dm}(q,\vk) = \frac{d\ln {\hat f^{dm}}_0}{d\ln q}  \; 
\psi(0,\vk)  -  \Psi_{dm}(0,q,\vk) \quad , \quad
F_{dm}(q,\vk) = - \frac{d\ln {\hat f^{dm}}_0}{d\ln q}  \;  
\frac{\partial{\phi}}{\partial \eta}(0,\vk) \; .
\ee
Similar equations hold for $ E_\nu(q,\vk) $ and $ F_\nu(q,\vk) $.

\medskip

Eqs.(\ref{iniPsi}) and (\ref{iniPsi3}) together with the integrals 
eqs.(\ref{sigma}) and (\ref{signu})
guarantee that eqs.(\ref{condS}) are fulfilled.

To the first order in $ \eta $ we obtain from eq.(\ref{bvy}) 
\be\label{der2ini}
\frac{\partial^2{\Psi_{dm}}}{\partial  \eta^2}(0,\vq,\vk) = \left( i \; \cq \cdot {\vk}\right)^2
\left[ \Psi_{dm}(0,q,\vk) - \frac{d\ln {\hat f^{dm}}_0}{d\ln q} \; \psi(0,\vk)\right]
+  i \; \cq \cdot {\vk} \; \frac{d\ln {\hat f^{dm}}_0}{d\ln q}
\left[\frac{\partial{\psi}}{\partial\eta }(0,\vk) + \frac{\partial{\phi}}{\partial \eta}(0,\vk) \right]
-\frac{d\ln {\hat f^{dm}}_0}{d\ln q}\frac{\partial^2{\phi}}{\partial \eta^2}(0,\vk) \; ,
\ee
and an analogous formula for the neutrino distribution function $ \Psi_\nu $.

\medskip

The knowledge of  the second derivative of the distribution functions with respect to
$ \eta $ at  $ \eta = 0 $ is necessary in order to compute the initial anisotropic stress
and the difference between $ \phi(0,\vk) $ and $ \psi(0,\vk) $ from eq.(\ref{ecpsi}).

\medskip

We compute the initial DM and neutrino density fluctuations from eqs.(\ref{t00}) and
(\ref{ecfinu2}), respectively
\be\label{delini}
\Delta_{dm}(0,\vk) = \frac{{\hat N_{dm}}\; g_{dm}}{2 \pi^2} 
\int_0^{\infty}dq \; q^3 \; {\hat f^{dm}}_0(q) \; 
 \Psi_{dm}(0,q,\vk)  \quad , \quad \Delta_{\nu}(0,\vk) = \frac{{\hat N_{\nu}}(0) \; g_{\nu}}{2 \pi^2} 
\int_0^{\infty}dq \; q^3 \; {\hat f^{\nu}}_0(q) \;  \Psi_\nu(0,q,\vk) \; .
\ee
Inserting this result in the linearized Einstein equations (\ref{eilin}) at  $ \eta = 0 $ gives
\be\label{cini}
\psi(0,\vk) = -\frac{4 \, \pi \; G \; {\eta^*}^2}{3 \, a_{eq}^2}
\left[\frac{{\hat N_{dm}}\; g_{dm}}{2 \pi^2} \int_0^{\infty}dq \; q^3 \; {\hat f^{dm}}_0(q) \; 
 \Psi_{dm}(0,q,\vk) + \frac{{\hat N}_{\nu}(0) \; g_{\nu}}{2 \pi^2} \int_0^{\infty}dq \; q^3 \; 
{\hat f^{\nu}}_0(q) \; \Psi_\nu(0,q,\vk) \right]- 2 \, R_\gamma(0) \; \Theta_0 (0, \vk) \; .
\ee
We compute the initial value of the anisotropic stress taking the $ \eta \to 0 $ limit
in eq.(\ref{sigma}) with the help of eq.(\ref{der2ini})
\bea
&&{\displaystyle \lim_{\eta \rightarrow 0}} \frac{\Sigma_{dm}(\eta, \vk)}{k^2 \; \eta^2} =
-\frac{{\hat N_{dm}}\; g_{dm}}{k^2} \int \frac{d^3q}{(2 \, \pi)^3} \; q \; 
P_2\left( {\check k} \cdot {\cq}\right) \;   {\hat f^{dm}}_0(q) \; 
\frac{\partial^2{\Psi_{dm}}}{\partial  \eta^2}(0,\vq,\vk)= \cr \cr
&& = {\hat N_{dm}}\; g_{dm} \int \frac{d^3q}{(2 \, \pi)^3} \; q \; 
P_2\left( {\check k} \cdot {\cq}\right) \; \left(\cq \cdot \check{k} \right)^2
\left[ \Psi_{dm}(0,q,\vk) - \frac{d\ln {\hat f^{dm}}_0}{d\ln q} \; \psi(0,\vk)\right]
\;   {\hat f^{dm}}_0(q) \; .
\eea
These integrals can be evaluated using eq.(\ref{delini}) and
$$
\int \frac{d\Omega({\breve q})}{4 \, \pi} \;\left(\cq \cdot \check{k} \right)^2 \; 
P_2\left( {\check k} \cdot {\cq}\right) = \frac2{15} \; ,
$$
with the final result
\bea\label{sigfin}
&&{\displaystyle \lim_{\eta \rightarrow 0}} \frac{\Sigma_{dm}(\eta, \vk)}{k^2 \; \eta^2} =
\frac2{15} \; \left[\Delta_{dm}(0,\vk) + \frac2{\pi^2} \; {\hat N}_{dm} \; g_{dm} \; \psi(0,\vk)
\; \int_0^{\infty}dq \; q^3 \; {\hat f^{dm}}_0(q)\right] \; , \cr \cr
&&{\displaystyle \lim_{\eta \rightarrow 0}} \frac{\Sigma_\nu(\eta, \vk)}{k^2 \; \eta^2} =
\frac2{15} \; \left[\Delta_\nu(0,\vk) + 4 \, R_\nu(0) \; \Omega_r \; \rho_c \; 
\psi(0,\vk) \right] \; .
\eea
Inserting this result in eq.(\ref{ecpsi}) gives the difference between the two gravitational 
potentials at the initial time 
\be\label{sigini}
\sigma(0, \vk) =  \phi(0, \vk) - \psi(0, \vk)  = \frac1{5 \, \rho_r} 
\left[\Delta_{dm}(0,\vk)+\Delta_{\nu}(0,\vk)\right] + \frac45\left[R_\nu(0) +
\frac{{\hat N_{dm}}\; g_{dm}}{2 \; \pi^2 \; \rho_r}\; 
\int_0^{\infty}dq \; q^3 \; {\hat f^{dm}}_0(q)\right]\psi(0,\vk)
\ee
where we used eqs.(\ref{ay}), (\ref{rnu}), (\ref{nnu})  and 
$$
\frac{4 \, \pi \; G \; {\eta^*}^2}{3 \, a_{eq}^2} = 
\frac1{2 \; \rho_r} \; .
$$

We see from eqs.(\ref{eilin}), (\ref{cini}) and (\ref{sigini})
that all dependence on $ \vk $ in the initial values of
$ \Delta_{dm}(0,\vk), \; \Delta_{\nu}(0,\vk), \; \Theta_0 (0, \vk) , \;  
\Psi_{dm}(0,q,\vk), \;  \Psi_\nu(0,q,\vk) $ and $ \sigma(0, \vk) $ can be taken proportional 
to $ \psi(0,\vk) $. We can therefore factor out $ \psi(0,\vk) $ from these initial values 
as 
\be\label{cbar}
\Psi_{dm}(0,q,\vk) = \psi(0,\vk) \; {\bar c}_{dm}^0(q) \quad , \quad 
\Psi_\nu(0,q,\vk) = \psi(0,\vk) \; {\bar c}_\nu^0(q) \; .
\ee
More generally, because the linear fluctuations evolve on an
homogeneous and isotropic cosmology, the linear evolution 
equations only depend on the modulus $ k $ (as we shall see 
explicitly in the next section), the dependence on the 
$ \vk $ directions keeps factorized for all times $ \eta $.
This is true for the distribution functions $ \Psi_{dm}(\eta, \vq, \vk) $
and $ \Psi_\nu(\eta, \vq, \vk) $ and for both gravitational potentials $ \psi $ and $ \phi $.

\medskip

Notice that from eq.(\ref{defnor}) 
\vskip -0.4 cm
$$
g_{dm} \; \frac{\hat N_{dm}}{2 \; \pi^2} \; T_d = \rho_{dm} \; \frac{T_d}{m} =
a_{eq} \; \frac{\rho_{dm}}{\xi_{dm}} \; ,
$$
and its neutrino counterpart eq.(\ref{nnu}).
\medskip

The initial gravitational potential $ \psi(0,\vk) $ is a Gaussian random field with variance
given by the primordial inflationary fluctuations \cite{biblia,dod,gor}
\be\label{fipot}
< \psi(0,\vk) \; \psi(0,\vk') > = \frac{P_\psi(k)}{(2 \, \pi)^3} \; \delta(\vk+\vk') \; ,
\ee
where we can use,
\be
P_\psi(k) = \frac{2 \, \pi^2}{k^3} \; \Delta_\psi^2(k) = \frac{8 \, \pi^2}9 \;
\frac{|\Delta_0 |^2}{k^3} \;\left(\frac{k}{k_0}\right)^{n_s-1} \quad , \quad
\Delta_\psi(k) = \frac23 \; \Delta_{\cal R}(k) \quad , \quad \Delta^2_{\cal R}(k)=
|\Delta_0 |^2 \; \left(\frac{k}{k_0}\right)^{n_s-1} \; .
\ee
The subscripts $ _\psi $ and $ _{\cal R} $ refer to the gravitational field 
and the scalar curvature, respectively.
$ | \Delta_0 | $ stands for the primordial power amplitude, $ n_s $ is the spectral index,
and $ k_0 $  is the pivot wavenumber \cite{WMAP,biblia}:
\be\label{ns}
|\Delta_0 | \simeq 4.94 \; 10^{-5} \quad , \quad n_s \simeq 0.964 \quad , \quad
k_0 = 2 \; {\rm Gpc}^{-1} \; .
\ee
The initial value of the gravitational potential $ \psi(0,\vk) $ can therefore
be written as
\be\label{fikp}
\psi(0,\vk) =  \frac{| \Delta_0 |}{3 \, \sqrt{\pi} \; k^\frac32} \; 
\left(\frac{k}{k_0}\right)^{\frac12(n_s-1)} \; g(\vk) \; ,
\ee
where $ g(\vk) $ is a Gaussian random field with unit variance
$$
< g(\vk) \; g^*(\vk') > = \delta(\vk-\vk') \; .
$$

\subsection{Physical magnitudes in dimensionless variables} 

From the analysis in the previous subsection we see that
it is convenient to define dimensionless density fluctuations 
and dimensionless anisotropic stress fluctuations
factoring out the initial gravitational potential $ \psi(0,\vk) $ 
in order to obtain quantities independent of the $ \vk $ direction:
\bea\label{defbar}
&&\Delta_{dm}(y,\vka) = {\bar \Delta}_{dm}(y,\alpha) \; \frac{g_{dm} \; 
{\hat N_{dm}} \; T_d}{2 \, \pi^2} \; \psi(0,\vk) \quad , \quad
\Delta_{\nu}(y,\vka) = {\bar \Delta}_{\nu}(y,\alpha) \; \frac{g_{\nu} \; 
{\hat N_{\nu}}(y) \; T^\nu_d}{2 \, \pi^2} \; \psi(0,\vk) \; , \cr \cr
&& \phi(y,\vka)=\psi(0,\vk) \; {\bar \phi}(y,\alpha) \quad , \quad 
\psi(y,\vka)=\psi(0,\vk) \; {\breve \psi}(y,\alpha) \quad {\rm and} \quad 
{\breve \psi}(0,\alpha) = 1 \; , \cr \cr
&& \sigma_{dm}(y,\vka) = \psi(0,\vk) \; {\bar \sigma}_{dm}(y,\alpha) 
\quad , \quad \sigma^\nu(y,\vka) = \psi(0,\vk) \; {\bar \sigma}^\nu(y,\alpha) 
\quad , \quad \sigma(y,\vka) = \psi(0,\vk) \; {\bar \sigma}(y,\alpha)
\; , \cr \cr
&& {\bar \sigma}(y,\alpha) = {\bar \phi}(y,\alpha) - {\breve \psi}(y,\alpha)
\quad , \quad {\bar \sigma}(0,\alpha)={\bar \phi}(0,\alpha)-1
\quad , \quad
\Theta_0 (y, \vka) =\psi(0,\vk) \; {\bar\Theta}_0 (y, \alpha) \; .
\eea
For ultrarelativistic neutrinos in dimensionless variables we have
[see eq.(\ref{deftzi})]:
\be \label{camnu}
E(\eta,q) \Rightarrow q = T_d^{\nu} \; Q \quad , \quad  
\varepsilon(y,Q) \Rightarrow Q \; .
\ee
The dimensionless density fluctuations are expressed in terms of the distribution functions as
\be \label{defDb}
{\bar \Delta}_{dm}(y,\kappa)= \int \frac{d^3Q}{4 \, \pi} \; \varepsilon(y,Q) \;
f_0^{dm}(Q) \; \frac{\Psi_{dm}(y, \vQ, \vka)}{\psi(0,\vka)} \quad , \quad
{\bar \Delta}_{\nu}(y,\kappa)=\int \frac{d^3Q}{4 \, \pi} \; Q \;
f_0^{\nu}(Q) \; \frac{\Psi_{\nu}(y, \vQ, \vka)}{\psi(0,\vka)} \quad ,
\ee
where we used eqs.(\ref{t00}), (\ref{ecfinu2}) and (\ref{defbar}).

\medskip

We find from eqs.(\ref{iniPsi}), (\ref{cbar}) and (\ref{defDb}), 
\be\label{I3bar}
{\bar \Delta}_{dm}(0,\kappa) = \int_0^{\infty} Q^3 \; dQ \; f_0^{dm}(Q) \; 
{\bar c}_{dm}^0(Q) \quad , \quad 
{\bar \Delta}_{\nu}(0,\kappa)= \int_0^{\infty}  Q^3 \; dQ \; f_0^{\nu}(Q) \; 
{\bar c}_{\nu}^0(Q) \; .
\ee
The customary DM and neutrino number density fluctuations are related to 
$ {\bar \Delta}_{dm}(y,\kappa) $ and $ {\bar \Delta}_{\nu}(y,\kappa) $ by
\be\label{deneu}
{\bar D}_{dm}(y,\kappa) = \frac1{4 \; I_3^{dm}} \;{\bar \Delta}_{dm}(y,\kappa)
\quad , \quad
{\bar N}_\nu(y,\kappa) = \frac1{4 \; I_3^\nu} \; {\bar \Delta}_{\nu}(y,\kappa) \; .
\ee

The linearized Einstein equations (\ref{ecfiy}) become for the dimensionless quantities
eq.(\ref{defbar}),
\be\label{ecpgsd}
\left[\left(1+{\cal R}_0(y)\right)  \left( \frac{d}{dy} + 1  \right) +
\frac13 \left(\kappa \; y \right)^2 \right]{\bar \phi}(y,\alpha) 
= [1+{\cal R}_0(y)] \;  {\bar \sigma}(y,\alpha) -\frac1{2 \, \xi_{dm}} \;  {\bar \Delta}_{dm}(y,\alpha)
-\frac{R_\nu(y)}{2 \, I^\nu_3}\;  {\bar \Delta}_\nu(y,\alpha)
-2 \,  R_\gamma(y) \; {\bar \Theta}_0(y,\alpha) \; .
\ee
From eqs.(\ref{sigini}) and (\ref{defbar}) the dimensionless density fluctuations and anisotropic stress 
fluctuations take as initial values,
\be\label{sbarini}
{\bar \sigma}(0,\alpha) = \frac15 \left[ \frac1{\xi_{dm}} \; {\bar \Delta}_{dm}(0,\alpha) 
+  \frac{R_\nu(0)}{I_3^\nu} \; {\bar \Delta}_{\nu}(0,\alpha) \right] + 
\frac45 \left[\frac{I_3^{dm}}{\xi_{dm}} + R_\nu(0)\right] \; .
\ee
Eqs.(\ref{ecpgsd}) and (\ref{sbarini}) suggest to introduce the quantities
\bea\label{dfDsom}
&& {\breve \Delta}(y,\alpha) \equiv \frac1{{\bar I}_\xi} 
\left[ \frac1{\xi_{dm}} \; {\bar \Delta}_{dm}(y,\alpha) 
+ \frac{R_\nu(y)}{I_3^\nu} \; {\bar \Delta}_{\nu}(y,\alpha) \right]
\quad , \quad {\breve \Delta}(0,\alpha) = 1 \; , \cr \cr
&& I_\xi \equiv \frac{I_3^{dm}}{\xi_{dm}} + R_\nu(0) \quad , \quad 
{\bar I}_\xi \equiv \frac{{\bar \Delta}_{dm}(0,\alpha)}{\xi_{dm}} + \frac{R_\nu(0)}{I_3^\nu} \;
{\bar \Delta}_{\nu}(0,\alpha)\; .
\eea
The relation between the initial values eq.(\ref{sbarini}) becomes,
\be\label{sbini}
5 \; {\bar \sigma}(0,\alpha) = 4 \, I_\xi + {\bar I}_\xi \quad {\rm and}
 \quad {\bar \phi}(0,\alpha) = 1 + \frac45 \, I_\xi + \frac15 \, {\bar I}_\xi \; .
\ee
The linearized Einstein equations (\ref{ecpgsd}) can be thus written in a 
more compact form
\be\label{beinlin}
\left[\left(1+{\cal R}_0(y)\right)  \left( \frac{d}{dy} + 1  \right) +
\frac13 \left(\kappa \; y \right)^2 \right]{\bar \phi}(y,\alpha) = 
[1+{\cal R}_0(y)] \;  {\bar \sigma}(y,\alpha) -
\frac12 \;  {\bar I}_\xi \;  {\breve \Delta}(y,\alpha)
-2 \, R_\gamma(y) \; {\bar \Theta}_0(y,\alpha) \; .
\ee
Eq.(\ref{beinlin}) at $ y = 0 $ gives the relation
\be\label{eli}
1 + \frac{I_3^{dm}}{\xi_{dm}}= -\frac12 \; {\bar I}_\xi -2 \,  R_\gamma(0) \; {\bar \Theta}_0(0,\alpha) \; .
\ee
where we used from eq.(\ref{R}) that $ {\cal R}_0(0) = I_3^{dm}/\xi_{dm} $.

\medskip

The initial number density fluctuations of photons 
$ {\bar \Theta}_0(0,\alpha)$, neutrinos $ {\bar N}_\nu(0,\alpha) $ and DM 
$ {\bar D}_{dm}(0,\alpha) $ are customary set equal to each other 
\cite{dod,mab,gor,weinb,pjep} which gives from eqs.(\ref{deneu}), (\ref{dfDsom}) and (\ref{eli})
\be\label{ibarxi}
1+ \frac{I_3^{dm}}{\xi_{dm}} =  -2 \; R_\nu(0) \; {\bar N}_\nu(0,\alpha)
-2 \; \frac{I_3^{dm}}{\xi_{dm}} \;
{\bar D}_{dm}(0,\alpha) -2 \,  R_\gamma(0) \; {\bar \Theta}_0(0,\alpha) \; ,
\ee
and therefore
\be\label{condI}
{\bar N}_\nu(0,\alpha)={\bar \Theta}_0(0,\alpha)={\bar D}_{dm}(0,\alpha)= 
-\frac12
\quad , \quad {\bar \Delta}_{dm}(0,\alpha) = -2 \; I_3^{dm} \quad , \quad 
{\bar \Delta}_{\nu}(0,\alpha) =  -2 \; I_3^{\nu} \; .
\ee
It follows in addition from eqs.(\ref{dfDsom}) and (\ref{condI}) that
\be\label{condI2}
 {\bar I}_\xi = -2 \; I_\xi \simeq - 2 \; R_\nu(0) \quad , \quad  
{\bar \sigma}(0,\alpha) = \frac25 \; I_\xi \simeq \frac25 \; R_\nu(0) \quad .
\ee
The approximation symbol $ \simeq $ here indicates that DM contributions 
to the initial data of the order $ 1/\xi_{dm} \ll 1 $ have been neglected.
As is known, DM is negligible in the RD era and its contributions to the 
initial data relative to the radiation contribution are of the order 
$ 1/\xi_{dm} $.

\medskip

Using eq.(\ref{condI2}) we can rewrite eq.(\ref{sbini}) as the relation 
between the two initial gravitational potentials 
$$
\phi(0,\vk) = \left[1 + \frac25 \; I_\xi \right]  \psi(0,\vk) \; .
$$
When corrections $ 1/\xi_{dm} $ are neglected 
this becomes a known relation \cite{mab,gor}
$$
\phi(0,\vk) \simeq \left[1 + \frac25 \; R_\nu(0) \right]  \psi(0,\vk) \; .
$$
In summary this yields for the initial gravitational potential
\be\label{fisini}
{\bar \phi}(0) \equiv {\bar \phi}(0,\alpha) = 
1 + \frac25 \; I_\xi \simeq 1 + \frac25 \; R_\nu(0) \; .
\ee
Eqs.(\ref{I3bar}), (\ref{condI}) and (\ref{condI2}) impose constraints on 
the functions 
$ {\bar c}_{dm}^0(Q) $ and $ {\bar c}_{\nu}^0(Q) $ defining the initial 
distribution functions.
We have to specify the initial functions $ {\bar c}_{dm}^0(Q) $ and 
$ {\bar c}_{\nu}^0(Q) $ to
completely define the initial data. There are two well motivated physical 
initial conditions. First, the thermal initial conditions (TIC)
(or thermal perturbation) \cite{primeros,mab,bdvs}, 
$$
T_d \rightarrow T_d \; \left[1+\frac{\delta T(\vk)}{T_d}\right] \; ,
$$
in which case $ {\bar c}_{dm}^0(Q) $ and $ {\bar c}_{\nu}^0(Q) $ are
proportional to $ d\ln f^{dm}_0/d\ln Q $ and  $ d\ln f^\nu_0/d\ln Q $,
respectively. Second, the Gilbert initial conditions (GIC) \cite{gil,bdvs} 
where $ {\bar c}_{dm}^0(Q) $ and $ {\bar c}_{\nu}^0(Q) $ are chosen to be 
constants. In order to fulfill eq.(\ref{condI}) we must choose for both DM 
and for neutrinos
\be\label{condini}
{\bar c}^0_{dm}(Q) =
\left\{\begin{array}{l} \displaystyle
\frac12 \;  \frac{d\ln f^{dm}_0}{d\ln Q} \quad 
{\rm for ~ thermal ~ initial ~ conditions ~ (TIC)} \; , \\ \\
-2 \quad {\rm for ~ Gilbert ~ initial ~ conditions ~ (GIC)} \quad .
\end{array} \right.
{\bar c}^0_\nu(Q) =
\left\{\begin{array}{l} \displaystyle
\frac12 \;  \frac{d\ln f^\nu_0}{d\ln Q} \quad 
{\rm for ~ TIC} \; , \\ \\
-2 \quad {\rm for ~ GIC} \quad .
\end{array} \right.
\ee
This completes the analysis of the initial conditions.

\section{The linear Boltzmann-Vlasov equation as a system of Volterra  integral 
 equations}\label{volt}

We recast in this section the linearized DM and neutrino B-V equations (\ref{bvy}) 
and (\ref{bvnu}) for $ \Psi(y, \vq, \vk) $ and  $ \Psi_\nu(y, \vq, \vk) $, coupled with 
the linearized Einstein's equation, as a system of linear integral equations of Volterra type.

\subsection{From the Boltzmann-Vlasov equations to the 
Volterra integral equations}

In the dimensionless variables eqs.(\ref{defQ})-(\ref{defnor})
the B-V equation (\ref{bvy}) takes the form
\be\label{bvsd}
\sqrt{1+ {\cal R}_0(y)} \; \frac{\partial{\Psi}_{dm}}{\partial y} + 
\frac{i \; \vQ \cdot {\vka}}{\varepsilon(y,Q)} \;
 \; \Psi_{dm}(y, \vQ, \vka) + \frac{d\ln f^{dm}_0}{d\ln Q}
\left[\sqrt{1+ {\cal R}_0(y)} \; \frac{\partial \phi}{\partial y}(y,\vka) - 
\frac{i \; \varepsilon(y,Q)}{Q^2} \; 
\vQ \cdot {\vka} \; \psi(y,\vka) \right] = 0  \; .
\ee
It is convenient to set 
\be\label{psi1} 
\Psi_{dm}(y, \vQ, \vka) = e^{-i \, \vka \cdot \vQ \; l(y,Q)/\xi_{dm}} 
\; \Psi_1(y, \vQ, \vka) =  e^{-i \, \va \cdot \vQ \; l(y,Q)/2} 
\; \Psi_1(y, \vQ, \vka)\; ,
\ee
where $ \va $ is related with $ \vka $ according to 
eqs.(\ref{defxi})-(\ref{alfs}) and
\be\label{rosd}
l(y,Q) \equiv \xi_{dm} \; \int_0^{y} \frac{dy'}{\varepsilon(y',Q) \; \sqrt{1+ {\cal R}_0(y')}}
= \int_0^{y} \frac{dy'}{\sqrt{\left[1+ {\cal R}_0(y')\right] \; \left[ y'^2 + \displaystyle 
\left(\displaystyle Q/\xi_{dm}\right)^2 \right]}} \; ,
\ee
the one-particle energy $ \varepsilon(y,Q) $ is defined by 
eq.(\ref{deftzi}). Notice that the free-streaming distance $ l(y,Q) $ 
depends on $ Q $ through the ratio $ Q/\xi_{dm} $. 
From eq.(\ref{R}) and the discussion after it, we can set from now on
$ {\cal R}_0(y) = y $ ignoring inessential 
$ 1/ \xi_{dm} $ or  $ 1/ \xi_{dm}^2 $ corrections.  (Except in sec. IV B of ref. \cite{dos}). 

\medskip

Since $ q/E(\eta,q) = Q/\varepsilon(y,Q) $ is the velocity of the DM particle at time 
 $ \eta $, its corresponding coordinate free-streaming length \cite{kt} is given by
\be\label{defro}
\lambda_{FS} = q \; \int_0^{\eta}\frac{d\eta'}{E(\eta',q)} =
\frac{\eta^*}{\xi_{dm}} \; Q \; l(y,Q) = \frac{l_{fs}}{2 \, \sqrt{I_4^{dm}}} \; Q \; l(y,Q)\; ,
\ee
where we used eqs.(\ref{ay}) and (\ref{rosd}). $ l_{fs} $ is given by eq.(\ref{deffs})
and sets the scale of the coordinate free-streaming length $ \lambda_{FS} $.

Inserting eq.(\ref{psi1}) into eq.(\ref{bv4}) yields for 
$ \Psi_1(y, \vQ, \vka) $ the equation
$$
\frac{\partial{\Psi_1}}{\partial y} = -\frac{d\ln f^{dm}_0}{d\ln Q}
 \; e^{+i \, \va \cdot \vQ \; l(y,Q)/2} \; \left[
\frac{\partial \phi}{\partial y} - \frac{i \; \varepsilon(y,Q)}{\sqrt{1+y} \; Q^2} 
\; \vka \cdot \vQ \; \psi(y,\vka) \right] \; .
$$
Integrating on $ y $ we obtain:
$$
\Psi_1(y, \vQ, \vka) = \Psi_1(0, \vQ, \vka) - \frac{d\ln f^{dm}_0}{d \ln Q}
\int_0^y d y' \;   e^{+i \, \va \cdot \vQ \; l(y',Q)/2} \;
\left[\frac{\partial \phi}{\partial y'} - \frac{i \; \varepsilon(y',Q)}{\sqrt{1+y'} \; Q^2} 
\; \vka \cdot \vQ \; \psi(y',\vka) \right] \; .
$$
Integrating the term $ \partial \phi/\partial y' $ by parts in $ y' $
and using eqs.(\ref{psi1}), we find for $ \Psi_{dm}(y, \vQ, \vka) $:
\bea\label{mul}
&& \Psi_{dm}(y, \vQ, \vka) = \psi(0,\vka) \; \left\{
{\bar c}^0_{dm}(Q) \; e^{-i \, \va \cdot \vQ \; l(y,Q)/2} +
\frac{d\ln f^{dm}_0}{d\ln Q} \left[ e^{-i \, \va \cdot \vQ \; l(y,Q)/2} \; {\bar \phi}(0,\alpha)
- {\bar \phi}(y,\alpha)\right] + \right. \cr \cr
&& \left. +i \; \frac{\vka \cdot \vQ}{Q^2} \; \frac{d\ln f^{dm}_0}{d\ln Q}
\int_0^y \frac{dy'}{\sqrt{1+y'}} \;   e^{+i \, \va \cdot \vQ \; [l(y',Q)-l(y,Q)]/2} \; 
\left(\left[\varepsilon(y',Q) + \frac{Q^2}{\varepsilon(y',Q)} \right] \;  {\bar \phi}(y',\alpha)
- \varepsilon(y',Q) \;  {\bar \sigma}(y',\alpha) \right)\right\} \; .
\eea  
Here we used eqs.(\ref{iniPsi}) and (\ref{cbar}) for the initial value $ \Psi_{dm}(0, \vQ, \vka) $.
Multiplying both sides of eq.(\ref{mul}) by $ \varepsilon(y,Q) \;  f^{dm}_0(Q) $, 
integrating over $ \vec Q $ and using eq.(\ref{defDb}) for
 the DM density fluctuations yields, 
\be\label{ecdel}
{\bar \Delta}_{dm}(y,\alpha)=  a(y,\alpha) + y \; \xi_{dm} \; 
b_{dm}(y) \; {\bar \phi}(y,\alpha)
 + \kappa \; \int_0^y \frac{dy'}{\sqrt{1+y'}} \;
\left[ N_\alpha(y,y')\; {\bar \phi}(y',\alpha)+
 N_\alpha^{\sigma}(y,y') \; {\bar \sigma}(y',\alpha) \right] \; ,
\ee
where we factored out the initial gravitational potential $ \psi(0,\vka) $ from the density
fluctuations according to eqs.(\ref{cbar})-(\ref{defbar}) in order to obtain a quantity 
independent of the directions of $ \va $:
\bea\label{varnor}
&& a(y,\alpha) \equiv  \int_0^{\infty}  Q^2 \; dQ \; \varepsilon(y,Q)\left[
f_0^{dm}(Q) \; {\bar c}_{dm}^0(Q) + {\bar \phi}(0) \; \frac{df_0^{dm}}{d\ln Q}\right]
j_0\left[\frac{\alpha}2 \, Q \, l(y,Q) \right] \label{asd3} \; ,\\ \cr \cr
&& 
 y \; \xi_{dm} \; b_{dm}(y) \equiv  \int_0^{\infty}\frac{Q^2 \; dQ}{\varepsilon(y,Q)}
\; f_0^{dm}(Q) \; \left[ 4 \, Q^2 + 3 \; (\xi_{dm} \; y)^2 \right] \label{dfb} \; ,\\ \cr \cr
&&  N_\alpha(y,y') =  \int_0^{\infty} Q^2 \; dQ \; \varepsilon(y,Q) \; 
\frac{df_0^{dm}}{dQ} \; j_1\left[\alpha \; l_Q(y,y') \right] \; 
\left[ \varepsilon(y',Q) +  \frac{Q^2}{\varepsilon(y',Q)}\right]  \label{byn}
\; ,\label{nucL2} \\ \cr \cr
&&  N_\alpha^{\sigma}(y,y') = -\int_0^{\infty} Q^2 \; dQ \; 
\frac{df_0^{dm}}{dQ} \; j_1\left[\alpha \; l_Q(y,y') \right] \;
\varepsilon(y,Q) \; \varepsilon(y',Q)   \; . \label{bynsi}
\eea
We used eqs.(\ref{intleg}) and (\ref{intxp2}), $ j_n(x) $ for $ 0 \leq n \leq 3 $ are spherical 
Bessel functions \cite{gr},
$$
l_Q(y,y') \equiv \frac12 \; Q \left[ l(y,Q)-l(y',Q) \right] = \frac{Q}2 \;
\int_{y'}^{y} \frac{dx}{\sqrt{(1+x) \; 
\left[ x^2 + \displaystyle \left(\displaystyle Q/\xi_{dm}\right)^2 \right]}} \; ,
$$
and we used the relation
\be\label{gorda}
\frac{4 \, \pi \; G \; {\eta^*}^2}{3 \, a_{eq}^2} \; \frac{g_{dm} \; N_{dm} \; T_d^4}{2 \, \pi^2} = 
\frac1{2 \; \xi_{dm}}  \; .
\ee
Notice from eq.(\ref{fisini}) that
$$ 
{\bar \phi}(0) \simeq 1 + \frac25 \; R_\nu(0) \; .
$$ 
The kernels $  N_\alpha(y,y') $ and $  N_\alpha^{\sigma}(y,y') $ only depend on the modulus of $ \va $ 
and {\bf not} on its direction since we consider linear fluctuations evolving on an homogeneous and 
isotropic cosmology.

\medskip

We derive now for $ {\bar \sigma}_{dm}(y,\alpha) $ an equation analogous to 
eq.(\ref{ecdel}). We first obtain from eqs.(\ref{sigma}), (\ref{ecpsi}) 
and (\ref{cbar})-(\ref{defbar}),
\be \label{defsi}
\psi(0,\va) \; {\bar \sigma}_{dm}(y,\alpha)= \frac{4 \, \pi \; G }{k^2 \; a^2(\eta)}
\; \Sigma_{dm}(\eta, \vk) = -\frac3{\xi_{dm} \; \kappa^2 \; y^2} \;
\int \frac{d^3Q}{4 \, \pi} \; \frac{Q^2}{\varepsilon(y,Q)} 
\; P_2\left({\check \kappa} \cdot {\cQ}\right)
\;  f^{dm}_0(Q) \; \Psi_{dm}(y, \vQ, \vka) \; .
\ee
We multiply eq.(\ref{mul}) by 
\be \label{peso}
 \frac{Q^2}{\varepsilon(y,Q)} \; P_2\left({\check \kappa} \cdot {\cQ}\right) \; f_0^{dm}(Q) \; \; ,
\ee
integrate over $ \vec Q $ and using eqs.(\ref{defDb}) and (\ref{defsi}) we find,
\be\label{ecsig}
\xi_{dm} \; {\bar \sigma}_{dm}(y,\alpha) =  
a^{\sigma}(y,\alpha) +  \kappa\;  \int_0^y  \frac{dy'}{\sqrt{1+y'}} \; \left[ U_\alpha(y,y') \; 
{\bar \phi}(y',\alpha) + U^{\sigma}_\alpha(y,y') \; {\bar \sigma}(y',\alpha)\right] \; ,
\ee
where,
\bea
&&  a^\sigma(y,\alpha) \equiv \frac3{\kappa^2 \; y^2}\int_0^{\infty}  
\frac{Q^4 \; dQ}{\varepsilon(y,Q)} \left[
f_0^{dm}(Q) \; {\bar c}_{dm}^0(Q) + {\bar \phi}(0) \; \frac{df_0^{dm}}{d\ln Q}\right] \; 
j_2\left[\frac{\alpha}2 \, Q \, l(y,Q) \right] \label{asd4} \; , \\ \cr \cr
&&  U_\alpha(y,y')= -\frac3{5 \, \kappa^2 \; y^2} 
\int_0^{\infty}  \frac{Q^4 \; dQ}{\varepsilon(y,Q)} \; \frac{df_0^{dm}}{dQ}
\; \left[ \varepsilon(y',Q) +  \frac{Q^2}{\varepsilon(y',Q)}\right]
\left\{ 2 \; j_1\left[\alpha \; l_Q(y,y') \right]
- 3 \; j_3\left[\alpha \; l_Q(y,y')\right] \right\}\; ,  \label{uay} \\ \cr \cr
&&  U^{\sigma}_\alpha(y,y') = \frac3{5 \, \kappa^2 \; y^2} 
\int_0^{\infty}  \frac{Q^4 \; dQ}{\varepsilon(y,Q)} \; \frac{df_0^{dm}}{dQ} \; \varepsilon(y',Q)
\left\{ 2 \; j_1\left[\alpha \; l_Q(y,y')\right] - 3 \; j_3\left[\alpha \; l_Q(y,y')\right]\right\}
\; . \label{uaysi} 
\eea
We used here eq.(\ref{intleg}) and (\ref{intxp2}). 

\medskip

Eqs.(\ref{ecdel}) and (\ref{ecsig}) form a system of Volterra equations 
\bea\label{ecvolz}
&& {\bar \Delta}_{dm}(y,\alpha)=   a(y,\alpha) + y \; \xi_{dm} \; 
b_{dm}(y) \; {\bar \phi}(y,\alpha)
 + \kappa \; \int_0^y \frac{dy'}{\sqrt{1+y'}} \;
\left[ N_\alpha(y,y')\; {\bar \phi}(y',\alpha)+
 N_\alpha^{\sigma}(y,y') \; {\bar \sigma}(y',\alpha) \right] \; \; , \cr \cr  
&& \xi_{dm} \; {\bar \sigma}_{dm}(y,\alpha) =   
 a^{\sigma}(y,\alpha) +  \kappa \int_0^y  \frac{dy'}{\sqrt{1+y'}} \; \left[ U_\alpha(y,y') \; 
{\bar \phi}(y',\alpha) + U^{\sigma}_\alpha(y,y') \; {\bar \sigma}(y',\alpha)\right] \; \; .
\eea
Notice that $ a(y,\alpha), \;   a^{\sigma}(y,\alpha), \;  {\bar \phi}(y,\alpha) , \;
{\bar \sigma}_{dm}(y,\alpha), \; {\bar \Delta}_{dm}(y,\alpha) $ and $ {\bar \sigma}(y,\alpha) $
{\bf only} depend on the modulus of $ \va $ and {\bf not} on the directions of  $ \va $.
The dependence on the $ \va $  directions comes from the 
initial power spectrum $ \psi(0,\vka) $ through the random field $ g(\vk) $
in eq.(\ref{fikp}) and turns to factor out, which
simplifies the resolution of the Volterra integral equations (\ref{ecvolz}). 
The factorization of the dependence on the $ \va $ directions
is possible because we consider linear fluctuations evolving on an
homogeneous and isotropic cosmology where all the evolution kernels
$ N_\alpha(y,y'), \; N^{\sigma}_\alpha(y,y'), \; U_\alpha(y,y') $
and $ U^{\sigma}_\alpha(y,y') $ are independent of the  $ \va $ directions. 

\medskip

The B-V distribution function as well as the coefficients in the B-V equation
depend on $ y, \; \va $ {\bf and} $ \vQ $. 
We integrate the distribution function over $ \vQ $ multiplied by appropriated weights.
The distribution function times $ \varepsilon(y,Q) $ produces the density eq.(\ref{defDb}) 
and the distribution function times the expression (\ref{peso}) produces
the anisotropic stress fluctuations eq.(\ref{defsi}). 
The density and the anisotropic stress fluctuations defined with such specific weights 
obey a {\bf closed} system of Volterra integral equations. 
Namely, no extra information on the $ \vQ $ dependence of the distribution 
functions is needed, which is a {\bf truly remarkable} fact.

\medskip

We derive below the Volterra integral equations for neutrinos eqs.(\ref{vurn})
similar to eqs.(\ref{ecvolz}) for DM.

\subsection{The pair of Volterra integral equations for DM and neutrinos}

The Volterra integral equations for neutrinos are obtained from eq.(\ref{bvnu}) 
following the same steps eqs.(\ref{psi1})-(\ref{uaysi}) which lead to 
the DM  Volterra integral equations (\ref{ecvolz}).
These Volterra integral equations for ultrarelativistic neutrinos are simpler than the
corresponding DM equations  
and follow from eqs.(\ref{asd3})-(\ref{uaysi}) making the following substitutions:
\bea\label{cambur}
&& \varepsilon(y,Q) \Rightarrow Q \quad , \quad  
\alpha \; l_Q(y,y') \Rightarrow \kappa \; r(y,y') \quad , \quad  
\alpha \; l(y,Q) \Rightarrow \frac{2 \; \kappa}{Q} \; r(y,0) \quad , \quad 
g_{dm} \; N_{dm} \Rightarrow  g^\nu  \; N^\nu(y) \quad , 
\\ \cr  && 
f_0^{dm}(Q) \Rightarrow  f_0^\nu(Q)  \quad , \quad 
{\bar \Delta}_{dm}(y,\alpha) \Rightarrow {\bar \Delta}_\nu(y,\alpha)  
\quad , \quad \xi_{dm} \Rightarrow \frac{I_3^\nu}{R_\nu(y)}
\quad , \quad 
\sigma_{dm}(y,\alpha)  \Rightarrow \sigma^\nu(y,\alpha) = 
\psi^\nu(y,\alpha) - \phi^\nu(y,\alpha) \; , \nonumber
\eea
where 
\vskip -0.5 cm
\be \label{defr}
r(y,y') \equiv 2 \, \left(\sqrt{1+y}-\sqrt{1+y'} \right) \quad , \quad
r(y,0) =  2 \, \left(\sqrt{1+y}-1\right)
\ee
and we used eq.(\ref{rosd}). [See also eq.(\ref{camnu})].

\medskip

Upon these changes the kernels $ N_\alpha(y,y'), \;   N_\alpha^{\sigma} (y,y'), \; 
U_\alpha(y,y'), \; U_\alpha^{\sigma} (y,y') $ in eqs.(\ref{byn})-(\ref{uaysi}) and the 
inhomogeneous terms $ a(y,\alpha) $ and $ a^{\sigma}(y,\alpha) $ in eq.(\ref{asd3}) 
simplify considerably. For ultrarelativistic neutrinos (ur) 
($ 0 < y < 34 \; m_{\nu}/0.05$ eV)
using eqs.(\ref{camnu})-(\ref{defr}) these kernels become:
\bea\label{rur}
&& N_\alpha(y,y') \buildrel{\rm ur ~ neutrinos} \over\Rightarrow N^{ur}_\alpha(y,y') \equiv
-8 \, I_3^{\nu} \; j_1\left[\kappa \; r(y,y')\right] \; , \cr \cr
&& U_\alpha(y,y')  \buildrel{\rm ur ~ neutrinos}\over\Rightarrow  U^{ur}_\alpha(y,y') \equiv
\frac{24 \, I_3^{\nu}}{5 \; \kappa^2 \; y^2} \; \left\{ 2 \; j_1\left[ \kappa \; r(y,y') \right] 
- 3 \;  j_3\left[ \kappa \; r(y,y') \right] \right\} \; , \\ \cr 
&& N_\alpha^{\sigma} (y,y')  \buildrel{\rm ur ~ neutrinos}\over\Rightarrow 
-\frac12 \;  N^{ur}_\alpha(y,y') \quad , \quad 
U_\alpha^{\sigma} (y,y')  \buildrel{\rm ur ~ neutrinos}\over\Rightarrow 
-\frac12 \;  U^{ur}_\alpha(y,y') \; , \\ \cr 
&& a(y,\alpha)  \buildrel{\rm ur ~ neutrinos}\over\Rightarrow 
a^{ur}(y,\alpha) \equiv -2 \, I_3^{\nu} \left[1 + 2 \; \bar \phi(0) \right] \;
j_0\left[ \kappa \; r(y,0)\right] \; \quad , \label{aur} \\ \cr 
&& a^{\sigma}(y,\alpha)  \buildrel{\rm ur ~ neutrinos}\over\Rightarrow  
a^{ur \; \sigma}(y,\alpha) \equiv -6 \, I_3^{\nu} \left[1 + 2 \; \bar \phi(0) \right] \;
\frac{j_2\left[\kappa \; r(y,0) \right]}{\kappa^2 \; y^2} 
\quad , \label{asur}
\eea
where we used eqs.(\ref{defin}), (\ref{I3bar}), (\ref{condI}), (\ref{asd3})-(\ref{bynsi}) 
and (\ref{asd4})-(\ref{uaysi}).

\medskip

In addition, when relevant the neutrinos are massless and using Table \ref{formu}, 
the coefficient of $ {\bar \phi}(y,\alpha) $ in eqs.(\ref{ecvolz}) for neutrinos becomes:
\vskip -0.6 cm
\be\label{defl}
y \; \xi_{dm} \; b_{dm}(y) \Rightarrow 4 \,  I_3^{\nu} \quad .
\ee
Therefore, making the changes eqs.(\ref{cambur})-(\ref{defl}) 
in eqs.(\ref{ecvolz}) yields the following Volterra integral equations 
for ultrarelativistic neutrinos 
\bea\label{vurn}
&& {\bar \Delta}^\nu(y,\alpha) =  a^{ur}(y,\alpha)
+4 \; I_3^{\nu} \; {\bar \phi}(y,\alpha) +\kappa \; \int_0^y \frac{dy'}{\sqrt{1+y'}} \; 
N^{ur}_\alpha(y,y') \left[  {\bar \phi}(y',\alpha) - 
\frac12 \; {\bar \sigma}(y',\alpha)\right] \; , \\ \cr \cr
&&  \frac{I_3^\nu}{R_\nu(y)} \; 
{\bar \sigma}^\nu(y,\alpha) = a^{ur \; \sigma}(y,\alpha)
+ \kappa \int_0^y  \frac{dy'}{\sqrt{1+y'}} \;  U^{ur}_\alpha(y,y') \; 
\left[{\bar \phi}(y',\alpha) - \frac12 \; {\bar \sigma}(y',\alpha)\right] \; . \label{vurn2}
\eea
Notice that the DM and neutrino Volterra integral equations eqs.(\ref{ecvolz}) and (\ref{vurn})
{\bf are coupled} to each other and to the linearized Einstein equations 
eq.(\ref{ecpgsd}) as well as to the hydrodynamic photon equations 
(\ref{hidro2})-(\ref{hidro}).

\medskip

It is possible to simplify the set of four Volterra integral equations
(\ref{ecvolz}) and (\ref{vurn}) into two Volterra equations.
Taking linear combinations of eqs.(\ref{ecvolz}) and (\ref{vurn})
we find for $ {\breve \Delta}(y,\alpha) $ [defined in eq.(\ref{dfDsom})]
and $ {\bar \sigma}(y,\alpha) $,
\bea\label{final}
&&{\breve \Delta}(y,\alpha) =  C(y,\alpha) + B_\xi(y)  \; {\bar \phi}(y,\alpha) +
\int_0^y dy' \left[G_\alpha(y,y') \; {\bar \phi}(y',\alpha) +
G^\sigma_\alpha(y,y') \; {\bar \sigma}(y',\alpha)\right] \; , \cr \cr\cr
&& {\bar \sigma}(y,\alpha) =  C^\sigma(y,\alpha) + \int_0^y dy'  \left[
I^\sigma_\alpha(y,y') \;  {\bar \sigma}(y',\alpha) + I_\alpha(y,y') \; {\bar \phi}(y',\alpha)
\right] \; ,
\eea
with the initial conditions eqs.(\ref{dfDsom}) and (\ref{condI2})
$$ 
{\breve \Delta}(0,\alpha) = 1 \quad , \quad {\bar \sigma}(0,\alpha) = \frac25 \; I_\xi 
\simeq \frac25 \; R_\nu(0)\quad .
$$ 
\vskip -0.2 cm
We have in eq.(\ref{final}) 
\bea\label{gorda2}
&& C(y,\alpha)  \equiv -\frac1{2 \, I_\xi} \; \left[ \frac{a(y,\alpha)}{\xi_{dm}} +\frac{R_\nu(y)}{I_3^\nu}\;
a^{ur}(y,\alpha)\right] \quad ,\quad  C^\sigma(y,\alpha)  \equiv \frac{a^\sigma(y,\alpha)}{\xi_{dm}} + 
\frac{R_\nu(y)}{I_3^\nu} \; a^{ur \; \sigma}(y,\alpha)\; , \label{calfa} \\ \cr 
&& B_\xi(y) \equiv -\frac1{2 \, I_\xi} \; \left[y \; b_{dm}(y) + 4 \, R_\nu(y)\right] \label{by}
\; , \cr \cr
&& G_\alpha(y,y') = -\frac{\kappa}{2 \, I_\xi \; \sqrt{1+y'} } \; \left[ \frac1{\xi_{dm}} \; N_\alpha(y,y') 
+ \frac{R_\nu(y)}{I_3^\nu} \; N^{ur}_\alpha(y,y') \right] 
\quad ,\label{galfa}  \\ \cr \cr   
&& G^\sigma_\alpha(y,y') =  -\frac{\kappa}{2 \, I_\xi \; \sqrt{1+y'}} \;
\left[\frac1{\xi_{dm}} \; N^\sigma_\alpha(y,y') 
- \frac{R_\nu(y)}{2 \, I_3^\nu} \; N^{ur}_\alpha(y,y') \right]  \label{gsalfa} \; , \\ \cr \cr 
&& I_\alpha(y,y') = \frac{\kappa}{\sqrt{1+y'}} \left[\frac1{\xi_{dm}} \; U_\alpha(y,y') + 
\frac{R_\nu(y)}{I_3^\nu} \; U^{ur}_\alpha(y,y')\right] \quad , \\ \cr \cr  
&& I^\sigma_\alpha(y,y')= \frac{\kappa}{\sqrt{1+y'}} \left[\frac1{\xi_{dm}} \; U^\sigma_\alpha(y,y') - 
\frac{R_\nu(y)}{2 \, I_3^\nu} \; U^{ur}_\alpha(y,y')\right] \; \label{defialfa}  \; .  
\eea
In eqs. (\ref{final}) we can use $ I_\xi \simeq R_\nu(0) $.

\medskip 

Notice that the $ G $ and $ I $ kernels 
in eqs.(\ref{galfa})-(\ref{defialfa}) result expressed as the sum 
of the DM contribution from the $ N $ and $ U $ kernels 
plus the (ultrarelativistic) neutrino 
contribution $ N^{ur}_\alpha(y,y') $ and $  U^{ur}_\alpha(y,y') $, respectively.
The inhomogeneous terms $ C(y,\alpha) $ and $ C^\sigma(y,\alpha) $ 
and the coefficient $ B_\xi(y) $ in eqs.(\ref{calfa})-(\ref{by}) turn also to be expressed 
as the sum of the DM plus the neutrino contributions.

\medskip 

In the MD era the neutrinos are negligible and its fraction $ R_\nu(y) $
becomes $ \ll 1 $ and can be neglected. Once neutrinos are negligible in the MD era,
the DM contribution to $ {\bar \sigma}(y,\alpha) $ from eqs.(\ref{final})-(\ref{defialfa}) 
is of the order $ 1/\xi_{dm} \ll 1 $ and the anisotropic stress becomes negligible.
This reduces the coupled Volterra integral equations (\ref{final})
in the MD era to a single Volterra integral equation for
$ {\breve \Delta}(y,\alpha) $ as we explicitly show in the accompanying paper \cite{dos}.

\medskip 

All functions in the inhomogeneous terms, coefficient and kernels in the the
Volterra equations (\ref{final}) are explicitly known from 
eqs.(\ref{gorda2})-(\ref{defialfa})
Therefore, eqs.(\ref{final}) plus the linearized Einstein equations 
(\ref{beinlin}) and the hydrodynamic photon
equations (\ref{hidro2})-(\ref{hidro}) provide a close system of equations 
determining $ {\breve \Delta}(y,\alpha), \; {\bar \phi}(y,\alpha) $ and  
$ {\bar \sigma}(y,\alpha) $. 
Once $ {\breve \Delta}(y,\alpha), \; {\bar \phi}(y,\alpha) $ and  
$ {\bar \sigma}(y,\alpha) $ are known we can
insert them in the r. h. s. of eqs.(\ref{ecvolz}) and (\ref{vurn}) to obtain
$ {\breve \Delta}_{dm}(y,\alpha), \;  {\bar \sigma}_{dm}(y,\alpha) , \;
{\breve \Delta}^\nu(y,\alpha) $ and $ {\bar \sigma}^\nu(y,\alpha) $, respectively.

\medskip 

We now set $ y = 0 $ in the system of Volterra equations (\ref{final}) to check their consistency. 
Taking into account eqs. (\ref{varnor})-(\ref{bynsi}), (\ref{asd4})-(\ref{uaysi}) 
and (\ref{gorda2})-(\ref{defialfa}) we obtain
\bea\label{muchas}
&& 
C(0,\alpha) =1 + 2 \; {\bar \phi}(0) \quad , \quad  
B_\xi(0) = - 2 \quad , \quad   
C^{\sigma}(0,\alpha)= -\frac25\left[1 + 2 \; {\bar \phi}(0)\right] \quad  ,
\cr \cr 
&& {\displaystyle \lim_{ y \rightarrow 0}} \; \int_0^y dy' \; G_\alpha(y,y')\; {\bar \phi}(y',\alpha) = 0
\quad , \quad  {\displaystyle \lim_{y \rightarrow 0}} \; \int_0^y dy' \; 
G_\alpha^{\sigma}(y,y') \; 
{\bar \sigma}(y',\alpha) = 0 \quad   , \label{epdel} \\ \cr 
&&  {\displaystyle \lim_{y \rightarrow 0}} 
\;  \int_0^y  dy' \; I_\alpha(y,y') \; {\bar \phi}(y',\alpha) 
= \frac85 \; I_\xi  \; {\bar \phi}(0) \quad , \quad 
{\displaystyle \lim_{y \rightarrow 0}} \; 
\int_0^y  dy' \; I^{\sigma}_\alpha(y,y') \;  {\bar \sigma}(y',\alpha) =  
- \frac8{25} \; I_\xi \; R_\nu(0) \quad   . \label{epsig}
\eea
Eqs.(\ref{final}) are identically satisfied at $ y = 0 $ due 
to eqs.(\ref{sbini}) and (\ref{muchas})-(\ref{epsig}).

\medskip

The system of Volterra equations (\ref{final}) is collisionless and it is 
therefore valid after both DM and neutrinos decoupled for 
$ y > y_d^\nu \simeq 0.5 \; 10^{-6} $ (see Table II). 
Since we are interested in adiabatic fluctuations which are regular 
solutions of eqs.(\ref{final}) at $ y = 0 $ we can start the evolution  
at $ y = 0 $ instead of $ y = y_d^\nu  \simeq 0.5 \; 10^{-6} $ with a 
negligible error.

For the DM particles, the range $ 0.5 \; 10^{-6} < y < 0.01 $ corresponds 
to the transition from ultrarelativistic to non-relativistic kinematics 
(see Table II).

\medskip

The density contrast $ \delta(y,\va) $ can be expressed in terms of the normalized
DM fluctuations $ {\bar \Delta}_{dm}(y,\alpha) $ from eqs.(\ref{Dd}) and (\ref{defbar}) as
\vskip -0.5 cm
\be\label{contra}
\delta(y,\va) = \frac1{\xi_{dm}} \; 
\frac{{\bar \Delta}_{dm}(y,\alpha)}{y+1} \; \psi(0,\va) \quad {\rm with} \quad 
\delta(0,\va) = -\frac{2 \, I_3^{dm}}{\xi_{dm}} \; \psi(0,\va) \; ,
\ee
where we used eq.(\ref{condI}) and $ \psi(0,\va) $ is given by the primordial fluctuations 
eq.(\ref{fikp}) 
and $ \xi_{dm} $ is given explicitly by eq.(\ref{forxi}).

\medskip

The integral equation (\ref{final}) supplemented by the fluid equations 
(\ref{hidro2})-(\ref{hidro})
for the photons and the linearized Einstein equations (\ref{ecpgsd}) 
{\bf provide a closed system of equations} to determine the DM, photon and neutrino density fluctuations.
This system of Volterra-type integral equations is valid for relativistic as well as 
non-relativistic particles propagating in the radiation and matter dominated eras.
This is the generalization of Gilbert's equation which is only valid
for non-relativistic particles in a matter dominated universe \cite{gil}. 

\medskip

We solve in an accompanying paper \cite{dos} the cosmological evolution of 
warm dark matter (WDM) density fluctuations presented here in the absence of 
neutrinos. In that case the anisotropic stress vanishes and the Volterra 
equations (\ref{final}) reduce to a single integral equation.

\acknowledgments

We are grateful to D. Boyanovsky and C. Destri for useful discussions.

\appendix

\section{The DM gravitational potential for large wavenumbers.}\label{ailam}

In sections \ref{bolvla} and \ref{volt} we found integrals of the type
\be\label{ila}
I_{\lambda}(y) = \int_0^y \frac{dx}{1+x} \; \left(\frac{1+y}{1+x} \; 
e^{x-y}\right)^{\lambda} \; f(x) \quad , \quad \lambda = \kappa^2/3 
\ee
The function $ \phi_{\lambda}(y) \equiv I_{\lambda}(y)/y $ solves
the first order differential equation
$$
\left[y(1+y) \; \frac{d}{dy} + 1 + y +  \lambda \; y^2 \right]\phi_{\lambda}(y) = f(y) \; ,
$$
which has the form of the linearized Einstein equations (\ref{ecpgsd}) and (\ref{beinlin}).

We derive here the asymptotic expansion of $ I_{\lambda}(y) $ in the limit where $ \lambda \gg 1 $.

\medskip

It is convenient to change the integration variable $ x $ in eq.(\ref{ila})
to $ s $ defined as
$$
s(x) \equiv \log\frac{1+x}{1+y}+y-x \quad , \quad s(y) = 0 \quad , \quad 
s(0) = y - \log(1+y) \; .
$$
The integral in eq.(\ref{ila}) becomes
$$
I_{\lambda}(y) = \int_0^{y- \log(1+y)} e^{-\lambda \; s} \; f[x(s)] \;
\frac{ds}{x(s)} \; .
$$
In the $ \lambda \gg 1 $ regime this integral is dominated by
the end-point of integration $ s = 0 $. Expanding $ f[x(s)]/x(s) $
around $ s = 0 $ and integrating term by term yields
\be\label{apr1}
I_{\lambda}(y) \buildrel{\lambda \gg 1}\over= \frac{f(y)}{\lambda \; y}
- \frac{1+y}{(\lambda \; y)^2} \; \left[\frac{df}{dy} - \frac{f(y)}{y} \right]
+ {\cal O}\left(\frac1{(\lambda \; y)^3}\right) \; .
\ee

\section{Angular Integrals}\label{angus}

We proceeded in sec. \ref{volt} to compute integrals over the directions of $ \vq $ with the help of the
partial wave expansion \cite{mf}
$$
e^{i \; \beta \; {\breve k} \cdot  {\breve q}} = \sum_{l=0}^\infty (2 \, l + 1) \; i^l \; j_l(\beta)
\; P_l\left({\breve k} \cdot  {\breve q}\right) \; .
$$
Integrating this expansion over the angles yields \cite{gr}
\be\label{intleg}
\int \frac{d\Omega({\breve q})}{4 \, \pi} \; e^{i \; \beta \; {\breve k} \cdot  {\breve q}} \;
 P_l\left({\breve k} \cdot  {\breve q}\right) = i^l \; j_l(\beta) \; .
\ee
We use in sec. \ref{volt} eq.(\ref{intleg}) for $ 0 \leq l \leq 3 $, the relations \cite{mf}
\be\label{recub}
j_0(x) = \frac{\sin x}{x} \quad , \quad j_{l+1}(x) = \frac{l}{x} \; j_l(x) - \frac{dj_l}{dx} 
\quad , \quad l \geq 0 
\ee
and the formulas for Legendre polynomials \cite{gr}
$$
P_0(x) = 1 \quad , \quad P_1(x) = x \quad , \quad P_{l+1}(x) = x \; \frac{2 \, l + 1}{l+1} \; P_l(x)
- \frac{l}{l+1} \; P_{l-1}(x) \quad , \quad l \geq 1 \; .
$$
It follows from these relations, in particular, that
\be\label{xp2}
x \; P_2(x) = \frac35 \;  P_3(x) + \frac25 \;  P_1(x) \; .
\ee
We get combining eqs.(\ref{intleg}) and (\ref{xp2}) ,
\be\label{intxp2}
\int \frac{d\Omega({\breve Q})}{4 \, \pi} \;   e^{+i \, \va \cdot \vQ \; [l(y',Q)-l(y,Q)]/2} \; 
 {\check \kappa} \cdot {\cQ} \; P_2\left( {\check \kappa} \cdot {\cQ}\right) =
-\frac{i}5 \; \left\{ 2 \; j_1\left[\alpha \; l_Q(y,y')\right] - 
3 \; j_3\left[\alpha \; l_Q(y,y')\right]\right\} \; .
\ee

\end{document}